\documentclass[a4paper,11pt]{article}
\usepackage{mathrsfs}
\usepackage{amssymb}
\usepackage{stmaryrd}
\usepackage{txfonts}
\usepackage[a4paper,left=3cm,right=3cm]{geometry}
\usepackage{graphicx}

\newcommand{\be}{\begin{equation}}
\newcommand{\ee}{\end{equation}}
\newcommand{\bea}{\begin{eqnarray}}
\newcommand{\eea}{\end{eqnarray}}

\makeatletter
\newcommand\figcaption{\def\@captype{figure}\caption}
\newcommand\tabcaption{\def\@captype{table}\caption}
\makeatother

\begin{document}
\setlength{\unitlength}{1.0mm}
\title{Muonium-Antimuonium Oscillations in an extended Minimal Supersymmetric Standard Model with right-handed neutrinos}
\author{Boyang Liu\thanks{liu115@physics.purdue.edu}\\
\small \emph{Department of Physics, Purdue University, West
Lafayette, IN 47906,USA}}
\date{}
\maketitle \normalsize
\begin{abstract}
The electron and muon number violating muonium-antimuonium
oscillation  process in an extended Minimal Supersymmetric Standard
Model is investigated. The Minimal Supersymmetric Standard Model is
modified by the inclusion of three right-handed neutrino
superfields. While the model allows the neutrino mass terms to mix
among the different generations, the sneutrino and slepton mass
terms have only intra-generation lepton number violation but not
inter-generation lepton number mixing. So doing, the
muonium-antimuonium conversion can then be used to constrain those
model parameters which avoid further constraint from the
$\mu\rightarrow e\gamma$ decay bounds. For a wide range of parameter
values, the contributions to the muonium-antimuonium oscillation
time scale are at least two orders of magnitude below the sensivity
of current experiments. However, if the ratio of the two Higgs field
VEVs, $\tan\beta$, is very small, there is a limited possibility
that the contributions are large enough for the present experimental
limit to provide an inequality relating $\tan\beta$ with the light
neutrino mass scale $m_\nu$ which is generated by see-saw mechanism.
The resultant lower bound on $\tan\beta$ as a function of $m_\nu$ is
more stringent than the analogous bounds arising from the muon and
electron anomalous magnetic moments as computed using this model.
\end{abstract}

\section{Introduction}
The time-dependent oscillation between two distinct levels or
particle species is an interesting quantum mechanical phenomenon
which has been widely studied in many physical systems varying from
a particle moving in a double-well potential of the ammonia molecule
to oscillations in the neutral $K^0-\bar K^0$ and $B^0-\bar B^0$
meson systems\cite{Battaglia}-\cite{Glashow}. It was suggested
roughly 50 years ago\cite{Pontecorvo} that there may be a
spontaneous conversion between muonium and antimuonium resulting in
an associated oscillation effect. Muonium $(M)$ is the Coulombic
bound state of an electron and an antimuon $(e^-\mu^+)$, while
antimuonium $(\bar M)$ is the Coulombic bound state of a positron
and a muon $(e^+\mu^-)$. Since it has no hadronic constituents,
muonium is an ideal place to test electroweak interactions. Of
particular interest is that such a muonium-antimuonium oscillation
is totally forbidden within the Standard Model because the process
violates the individul electron and muon number conservation laws by
two units. Hence, its observation will be a clear signal of physics
beyond the Standard Model. Since the initial suggestion,
experimental searches have been conducted\cite{M-A
expts}-\cite{Willmann} and a variety of theoretical models have been
proposed which can give rise to such a muonium-antimuonium
conversion. These include interactions which can be mediated by (a)
a doubly charged Higgs boson $\Delta^{++}$\cite{Halprin,Herczeg},
which is contained in a left-right symmetric model, (b) massive
Majorana neutrinos\cite{Clark&Love, Kim, Liu}, or (c) the
$\tau$-sneutrino in an R-parity violation supersymmetric
model\cite{Halprin and Masiero}.

In this paper we consider the muonium-antimuonium oscillation
process in the Minimal Supersymmetric Standard Model extended by the
inclusion of three right-handed neutrino superfields. While the
neutrino mass terms can mix inter-generationally, we allow only
intra-generation lepton number violation but not inter-generation
lepton number mixing for the sneutrino and slepton mass terms. In
this model, there are intermediate states which can contribute to
the muonium-antimuonium oscillation process but not to the
$\mu\rightarrow e\gamma$ decay. Therefore, the experimental limit on
muonium- antimuonium oscillations can be used to constrain those
model parameters which are not constrained by the $\mu\rightarrow
e\gamma$ decay bounds. In order for there to be  a nontrivial mixing
between the muonium and antimuonium, the individual electron and
muon number conservation must be violated by two units. Such a
situation will result provided that the neutrinos are massive
Majorana particles or the mass diagonal sneutrinos are lepton number
violating scalar particles.

\section{The extended Minimal Supersymmetric Standard Model}
The Minimal Supersymmetric Standard Model (MSSM) is the
supersymmetric extension of the (2 scalar doublet) Standard Model
with the minimal particle content\cite{Martin}. For each particle,
there is a superpartner with the same internal quantum numbers, but
with spin that differs by half a unit. Tab. 1 lists all the chiral
supermultiplets needed for MSSM,
\begin{center}
\begin{tabular}{|c|c||c|c|c|}\hline
\multicolumn{2}{|c||}{\textbf{Names}} & \textbf{Spin} 0 & $
\textbf{Spin} \frac{1}{2}$ & $\textbf{SU(3)
}_C,\textbf{SU(2)}_L,\textbf{U(1)}_Y$\\ \hline\hline
squarks & Q & $(\tilde u_L~~\tilde d_L)$ & $(u_{L\alpha}~~d_{L\alpha})$ & $(3, 2, \frac{1}{6})$\\
and quarks & $U^c$ & ${\tilde u}_R^\ast$ & $\bar u_R^{\dot{\alpha}}$ & $(\bar 3, 1, -\frac{2}{3})$\\
($\times 3$ families) & $D^c$ & ${\tilde d}_R^\ast$ & $\bar
d_R^{\dot{\alpha}}$ & $(\bar 3, 1, \frac{1}{3})$\\ \hline
sleptons, leptons & L & $(\tilde\nu_L ~~\tilde e_L)$ & $ (\nu_{L\alpha}~~e_{L\alpha}) $ & $(1, 2, -\frac{1}{2})$ \\
($\times 3$ families)& $ E^c $ & $\tilde e^\ast_R $ & $\bar
e_R^{\dot{\alpha}} $ & $ (1, 1, 1)$\\ \hline
Higgs, higgsinos &$ H_T$ & $(h_T^+~~h^0_T)$ & $(\tilde h^+_{T\alpha} ~~\tilde h^0_{T\alpha})$ & $(1, 2, \frac{1}{2})$\\
&$ H_B$ & $(h_B^0~~h^-_B)$ & $(\tilde h^0_{B\alpha} ~~\tilde
h^-_{B\alpha})$ & $(1, 2, -\frac{1}{2})$\\\hline
\end{tabular}
\end{center}
\begin{center}
\tabcaption{Chiral supermultiplets of the MSSM}
\end{center}
while Tab. 2 summarizes the gauge supermutiplets of MSSM.
\begin{center}
\begin{tabular}{|c||c|c|c|}\hline
\textbf{Names} & \textbf{Spin} $\frac{1}{2}$ &  \textbf{Spin} 1
& $\textbf{SU(3) }_C,\textbf{SU(2)}_L,\textbf{U(1)}_Y$\\
\hline\hline Gluino, gluon &$ \tilde g_\alpha$&$g_\mu$&$(8, 1, 0)$\\
\hline winos, W bosons & $\tilde W^\pm_\alpha~~\tilde W^0_\alpha$ &
$W^\pm_\mu~~W^0_\mu$
&(1, 3, 0)\\ \hline bino, B boson & $\tilde B_\alpha$ & $B_\mu$ & $(1, 1, 0)$\\
\hline
\end{tabular}
\end{center}
\begin{center}
\tabcaption{Gauge supermultiplets in the MSSM}
\end{center}
In the above two tables, the dotted and undotted indices, $\alpha$,
$\dot\alpha$, indicate 2-component Weyl spinor fields. In the
subsequent analysis, we will recast all the spin $\frac{1}{2}$
fields as 4-component Dirac spinor fields, which will be represented
using the same symbols, but without the dotted and undotted
``$\alpha$"s. For example, $ \nu_{L\alpha}$ is the Weyl
representation of the left-handed neutrino field, while
$\nu_L=\left(\begin{array}{c}\nu_{L\alpha}\\\bar\nu_L^{\dot\alpha}\end{array}\right)$
is the Dirac field.

In order to implement the see-saw mechanism\cite{seesaw} for
neutrino masses, we consider an extension of the MSSM, where one
adds three additional gauge singlet chiral superfields $N^c_i$
(i=$e,\mu,\tau$ denotes the generation), whose $\theta$-component is
a right-handed neutrino field, \be N^c_i=\tilde
\nu^\ast_{iR}(y)+\sqrt 2\theta^\alpha
\nu_R(y)_\alpha+\theta^\alpha\theta_\alpha F_{N^c_i}(y),\ee where
\be
y^\mu=x^\mu+i\theta^\alpha\sigma^\mu_{\alpha\dot\alpha}\bar\theta^{\dot\alpha}.
\ee These $SU(3)\times SU(2)_L\times U(1)$ singlet superfields are
coupled to other MSSM superfields via the superpotential. We employ
the most general R-parity conserving renormalizable superpotential
so that the superpotential is \be W=-\mu \epsilon_{ab} H^a_B
H^b_T+\lambda_i
\epsilon_{ab}E^c_iL^a_iH^b_B+\lambda^\prime_{ij}\epsilon_{ab}H^a_TL^b_iN^c_j+\frac{1}{2}M^{ij}_R
N^c_iN^c_j,\ee while the relevant soft supersymmetry breaking terms
are \bea {\cal L}^{EMSSM}_{soft}=&&-(m^{ij}_{\tilde L})^2\Big(\tilde
\nu^\ast_{iL}\tilde
\nu_{jL}+\tilde\ell_{iL}^\ast\tilde\ell_{jL}\Big)-(m^{ij}_{\tilde
R})^2\tilde \ell^\ast_{iR}\tilde\ell_{jR}-(m^{ij}_N)^2\tilde
\nu^\ast_{iR}\tilde \nu_{jR}\cr &&-\Big(\lambda^\prime_{ij} A_{ij}
h^0_T\tilde\nu_{iL}\tilde \nu^\ast_{jR}+M^{ij}_R
B_{ij}\tilde\nu_{iR}\tilde\nu_{jR}+\lambda_i
C_{ii}h^0_B\tilde\ell_{iL}\tilde\ell^\ast_{iR}+
C_{ij}h^0_B\tilde\ell_{iL}\tilde\ell^\ast_{jR}+H.C.\Big).\eea

The interaction terms that contribute to the muonium-antimuonium
oscillation and the electron and muon anomalous magnetic moments can
be extracted from the Lagrangian of  this extended Minimal
Supersymmetric Standard Model (EMSSM) as \be {\cal
L}^W_{int}=-\frac{g_2}{\sqrt 2}\Big(W^{-\mu}\bar \ell_{iL}\gamma_\mu
\nu_{iL}+W^{+\mu}\bar \nu_{iL}\gamma_\mu \ell_{iL}\Big),\ee \be
{\cal L}^{\widetilde W^-}_{int}= -ig_2\Big(\bar \ell_{iL}~\widetilde
W^-~\widetilde\nu_{iL}-{\widetilde
\nu}^\ast_{iL}~\overline{\widetilde W^-}~\ell_{iL}\Big),\ee \be
{\cal L}^{\widetilde W^0}_{int}=\frac{g_2i}{\sqrt 2}\Big(\bar
\ell_{iL}~\widetilde W^0 ~\widetilde
\ell_{iL}-\widetilde\ell^\ast_{iL}~\overline{\widetilde
W^0}~\ell_{iL}\Big),\ee \be {\cal L}^{\widetilde
B}_{int}=\frac{g_1i}{\sqrt 2}\Big(\bar \ell_{iL}~\widetilde B
~\widetilde \ell_{iL}-\widetilde\ell^\ast_{iL}~\overline{\widetilde
B}~\ell_{iL}\Big)+\sqrt 2 g_1i\Big(\bar \ell_{iR}~\widetilde B
~\widetilde \ell_{iR}-\widetilde\ell^\ast_{iR}~\overline{\widetilde
B}~\ell_{iR}\Big),\ee \be {\cal L}^{\widetilde
h^-_B}_{int}=\frac{m_i}{V_B}\Big(\bar
\ell_{iR}~\widetilde{h^-_B}~\widetilde\nu_{iL}+\overline{\widetilde{h^-_B}}~\ell_{iR}~\widetilde\nu^\ast_{iL}\Big)
+\Big(\frac{m_D^{ij}}{V_T}\bar
\ell_{iL}~\widetilde{h^-_B}~\widetilde\nu_{iR}+\frac{(m_D^{ij})^\ast}{V_T}\overline{\widetilde{h^-_B}}~\ell_{iL}~\widetilde\nu^\ast_{iR}\Big),\ee
\be {\cal L}^{\widetilde h^0_B}_{int}=-\frac{m_i}{V_B}\Big(\bar
\ell_{iL}~\widetilde h^0_B~\widetilde\ell_{iR}+\widetilde
\ell^\ast_{iR}~\overline{\widetilde
h^0_B}~\ell_{iL}\Big)-\frac{m_i}{V_B}\Big(\bar \ell_{iR}~\widetilde
h^0_B~\widetilde\ell_{iL}+\widetilde
\ell^\ast_{iL}~\overline{\widetilde h^0_B}~\ell_{iR}\Big).\ee In the
above equations, all the spin $\frac{1}{2}$ fields are Dirac spinor
fields. In particular, note that the field $\widetilde h^-_B$ has
the Weyl field decomposition \bea \widetilde
h^-_B=\left(\begin{array}{c} \widetilde h^-_{B\alpha}
\\ \overline{\widetilde h^{+}_T}^{\dot\alpha} \end{array} \right).\eea
The parameters $V_B$ and $V_T$ are the vacuum expectation values of
the two Higgs fields: $ <h^0_B>=V_B$ and $<h^0_T>=V_T.$ These VEVs
are related to the known mass of the $W$ boson and the electroweak
gauge couplings as \be
V_B^2+V_T^2=V^2=\frac{2M^2_W}{g^2_2}\approx(174GeV)^2,\ee while the
ratio of the VEVs is traditionally written as\be
\tan\beta\equiv\frac{V_T}{V_B}.\ee In the above,
$m_D^{ij}=\lambda^\prime_{ij} V_T$ is the Dirac mass matrix of
neutrinos and $m_i$ are the lepton masses. Since the masses of
electron and muon are small, the terms which have couplings
proportional to $m_i/V_B$ in interactions (9) and (10) are severely
suppressed and will be ignored in the subsequent analysis.

The neutrino mass term can be extracted from the superpotential
terms $\lambda^\prime_{ij}\epsilon_{ab}H^a_TL^b_iN^c_j$ and
$\frac{1}{2}M^{ij}_R N^c_iN^c_j$ as\be {\cal L}^{\nu}_{mass}=
-\frac{1}{2}\left(\begin{array}{c}\overline{(\nu_{L})^c}
~~\overline{\nu_{R}}\end{array}\right)
\left(\begin{array}{c}0~~~~~~ m^T_D\\m_D~~~~~~M_R\end{array}\right)\left(\begin{array}{c}\nu_{L}\\
(\nu_{R})^c\end{array}\right)+H.C.,\ee where\be
\nu_{L}=\left(\begin{array}{c}\nu_{eL}\\\nu_{\mu L}\\\nu_{\tau
L}\end{array}\right),~~~~~~\nu_{R}=\left(\begin{array}{c}\nu_{eR}\\\nu_{\mu
R}\\\nu_{\tau R}\end{array}\right).\ee Note that the upper left
$3\times3$ block in the neutrino mass matrix is zero. This block
matrix involves only left-handed neutrinos and in our EMSSM its
generation requires a nonrenormalizable superpotential term.
Consequently we ignore this term. For three generations of
neutrinos, the six mass eigenvalues, $m_{\nu a}$, are obtained from
the diagonalization of the $6\times 6$ matrix\be
M^{\nu}=\left(\begin{array}{c}0~~~~~~
m^T_D\\m_D~~~~~~M_R\end{array}\right).\ee Since $M^{\nu}$ is
symmetric, it can be diagonalized by a single unitary $6\times6$
matrix, $V$, as\be M^{\nu}_{diag}=V^{T}M^{\nu}V.\ee This
diagonalization is implemented via the basis change as following \be
\left(\begin{array}{c}\nu_{L}\\(\nu_{R})^c\end{array}\right)
=\left(\begin{array}{c}\nu_{eL}\\\nu_{\mu L}\\\nu_{\tau
L}\\(\nu_{eR})^c\\(\nu_{\mu R})^c\\(\nu_{\tau
R})^c\end{array}\right)=V
\left(\begin{array}{c}\nu_{1}\\\nu_{2}\\\nu_{3}\\(\nu_{4})^c\\(\nu_{5})^c\\(\nu_{6})^c\end{array}\right),
~~~~~~~ \left(\begin{array}{c}(\nu_{L})^c\\
\nu_{R}\end{array}\right)=\left(\begin{array}{c}(\nu_{eL})^c\\(\nu_{\mu L})^c\\(\nu_{\tau L})^c\\
\nu_{eR}\\\nu_{\mu R}\\\nu_{\tau R}\end{array}\right)=V^{\ast}\left(\begin{array}{c}(\nu_{1})^c\\(\nu_{2})^c\\(\nu_{3})^c\\
\nu_{4}\\\nu_{5}\\\nu_{6}\end{array}\right).\ee The neutrino mass
term then takes the form\be {\cal
L}^{\nu}_{mass}=-\frac{1}{2}\sum_{a=1}^6m_{\nu
a}[\nu^T_{a}C\nu_{a}+\overline{\nu_{a}}C\overline{\nu_{a}^T}]=
-\sum_{a=1}^6m_{\nu a}\overline{\nu_{a}}\nu_{a}, \ee where $m_{\nu
a}$ are the Majorana neutrino masses.

Since a nonzero Majorana mass matrix $M^{ij}_R$ does not require
$SU(2)_L\times U(1)$ symmetry breaking, it's naturally characterized
by a much larger scale, $M_R$, than $m_D$, which is the scale of the
Dirac mass matrix $m^{ij}_D$ whose nontrivial value does require
$SU(2)_L\times U(1)$ symmetry breaking. So doing, one finds on
diagonalization of the $6\times6$ neutrino mass matrix that the
three eigenvalues are crudely given by\be m_{\nu
a}\sim\frac{m_D^2}{M_R}\ll m_D,~~~~~a=1, 2, 3, \ee while the other
three eigenvalues are roughtly\be m_{\nu a}\sim M_R, ~~~~~~a=4, 5,
6.\ee This constitutes the so called see-saw mechanism\cite{seesaw}
and provides a natural explanation of the smallness of the three
light neutrino masses. Moreover, the elements of the mixing matrix
are characterized by an $m_D/M_R$ dependence \bea &&V_{ab}\sim
\mathcal {O}(1), ~~~~a,b=1,2,3,\cr &&V_{ab}\sim\mathcal{O}(1),
~~~~a,b=4,5,6,\cr &&V_{ab}\sim
V_{ba}\sim\mathcal{O}(\frac{m_D}{M_R}), ~~~~a=1,2,3, b=4,5,6. \eea

In order to obtain the sneutrino masses, it's convenient to define
$\tilde\nu_{iL}=\frac{1}{\sqrt 2}(\tilde\nu_{iL1}+i\tilde\nu_{iL2})$
and $\tilde\nu_{iR}=\frac{1}{\sqrt
2}(\tilde\nu_{iR1}+i\tilde\nu_{iR2})$. Then, the sneutrino-squared
mass matrix separates into CP-even and CP-odd blocks\cite{Grossman},
\bea {\cal L}^{\tilde
\nu}_{mass}=&&\sum_{i,j=e,\mu,\tau}\frac{1}{2}\left(\begin{array}{cc}\phi_1^i
 & \phi_2^i\end{array}\right){\mathcal M}^2_{\tilde \nu_{ij}}
\left(\begin{array}{c}\phi_1^j\\
\phi_2^j\end{array}\right)\cr=&&\sum_{i,j=e,\mu,\tau}\frac{1}{2}\left(\begin{array}{cc}\phi_1^i
 & \phi_2^i\end{array}\right)
\left(\begin{array}{cc}{\mathcal M}^2_{\tilde \nu_{ij}+} & 0\\0 &
{\mathcal M}^2_{\tilde \nu_{ij}-}\end{array}\right)
\left(\begin{array}{c}\phi_1^j\\
\phi_2^j\end{array}\right),\eea where
$\phi^i_a\equiv(\tilde\nu_{iLa}~~~\tilde\nu_{iRa})$ and ${\mathcal
M}^2_{\tilde \nu_{ij}\pm}$ consist of the following $2\times2$
blocks:\be {\mathcal M}^2_{\tilde \nu_{ij}\pm}=\left(
\begin{array}{cc}(m^{ij}_{\tilde L})^2+\frac{1}{2}m^2_Z
\cos{2\beta}+(m^{ij}_D)^2 & m^{ij}_D(A_{ij}-\mu\cot{\beta}\pm
M^{ij}_R)\\m^{ij}_D(A_{ij}-\mu\cot{\beta}\pm M^{ij}_R) &
(M^{ij}_R)^2+(m^{ij}_D)^2+(m^{ij}_{\tilde N})^2\pm2B_{ij}
M^{ij}_R\end{array}\right),\ee with $A_{ij}$ and $B_{ij}$ are SUSY
breaking parameters (cf. Eq.(4)). Since we allow only
intra-generation lepton number violation but not inter-generation
lepton number mixing for the supersymmetric partners, we can arrange
the parameters in matrices  (24) so that ${\mathcal M}^2_{\tilde
\nu_{ij}\pm}=0$ for $i\neq j$. So doing, the sneutrino mass term
simplifies into three $4\times4$ matrices for three generations \bea
{\cal L}^{\tilde
\nu}_{mass}=&&\sum_{i=e,\mu,\tau}\frac{1}{2}\left(\begin{array}{cc}\phi_1^i
 & \phi_2^i\end{array}\right)
\left(\begin{array}{cc}{\mathcal M}^2_{\tilde \nu_{ii}+} & 0\\0 &
{\mathcal M}^2_{\tilde \nu_{ii}-}\end{array}\right)
\left(\begin{array}{c}\phi_1^i\\
\phi_2^i\end{array}\right).\eea The sneutrino mass matrix ${\mathcal
M}^2_{\tilde \nu_{ii}}$ is real and symmetric, so it can be
diagonalized by a real orthogonal $4\times4$ matrix, $U^i$, as \be
{\mathcal M}^2_{\tilde\nu_i diag}=U^{iT} {\mathcal M}^2_{\tilde
\nu_{ii}} U^i,\ee where $U^i$ is in a form as \be U^i=\left(
\begin{array}{cc}U^i_+ & 0\\0 & U_-^i\end{array}\right).\ee This
diagonalization is implemented via the basis change on $\phi^i_1$
and $\phi^i_2$
\be\left(\begin{array}{c}\phi_1^i\\
\phi_2^i\end{array}\right)=\left(\begin{array}{c}\tilde\nu_{iL1}\\\tilde\nu_{iR1}\\\tilde\nu_{iL2}\\\tilde\nu_{iR2}
\end{array}\right)=U^i\left(\begin{array}{c}\tilde\nu_{i1}\\\tilde\nu_{i2}\\\tilde\nu_{i3}\\\tilde\nu_{i4}
\end{array}\right),\ee where $\tilde \nu_{ia}$ are all real.
Then the sneutrino mass term takes the form \be {\cal
L}^{\tilde\nu_i}_{mass}=-\frac{1}{2}\sum^4_{a=1}
m^{\tilde\nu_i}_a\tilde\nu_{ia}\tilde\nu_{ia},\ee where
$m^{\tilde\nu_i}_a$ are the sneutrino mass eigenvalues.

In the following derivation we assume that $M^{ii}_R$ is the largest
mass parameter. Then, to the first order in $1/M^{ii}_R$, the two
light mass eigenvalues are roughly\bea
&&m^2_{\tilde\nu_{i1}}\approx(m^{ii}_{\tilde
L})^2+\frac{1}{2}m^2_Z\cos{2\beta}-\frac{2(m^{ii}_D)^2
(A_{ii}-\mu\cot\beta -B_{ii})}{M^{ii}_R},\cr
&&m^2_{\tilde\nu_{i3}}\approx(m^{ii}_{\tilde
L})^2+\frac{1}{2}m^2_Z\cos{2\beta}+\frac{2(m^{ii}_D)^2
(A_{ii}-\mu\cot\beta -B_{ii})}{M^{ii}_R},\eea while the two heavy
mass eigenvalues are\bea
&&m^2_{\tilde\nu_{i2}}\approx(M^{ii}_R)^2+2B_{ii}M^{ii}_R,\cr
&&m^2_{\tilde\nu_{i4}}\approx(M^{ii}_R)^2-2B_{ii}M^{ii}_R. \eea

To avoid excessive complication in our calculations, we expand $U^i$
in powers of the matrix parameter $\xi_i=\frac{m^{ii}_D}{M^{ii}_R}$.
The form of $U$ to first order of $\xi_i$ is\bea U^i=&&\left(
\begin{array}{cc}U^i_+ & 0\\0 & U^i_-\end{array}\right)\cr=&&\left(
\begin{array}{cc}\left(\begin{array}{cc}1 & \xi_i\\ -\xi_i & 1\end{array}\right) & 0\\0 &
\left(\begin{array}{cc}1 & -\xi_i\\ \xi_i &
1\end{array}\right)\end{array}\right).\eea

The slepton mass term is given by \bea {\cal L}^{\tilde
\ell}_{mass}=\sum_{i,j=e,\mu,\tau}\left(\begin{array}{cc}\tilde\ell_{iL}^\ast
&
\tilde\ell^\ast_{iR}\end{array}\right)\left(\begin{array}{cc}(m^{LL}_{\tilde\ell_{ij}})^2
& (m^{LR}_{\tilde\ell_{ij}})^2\\(m^{LR}_{\tilde\ell_{ij}})^2 &
(m^{RR}_{\tilde\ell_{ij}})^2
\end{array}\right)\left(\begin{array}{c}\tilde\ell_{jL}\\ \tilde\ell_{jR}\end{array}\right), \eea where\be
(m^{LL}_{\tilde\ell_{ij}})^2=(m^{ij}_{\tilde
L})^2+m_Z^2\cos{2\beta}\Big(\sin^2{\theta_W}-\frac{1}{2}\Big),\ee
\be(m^{LR}_{\tilde\ell_{ii}})^2=\lambda_i\mu V_T+\lambda_iC_{ii}
V_B,~~~~(m^{LR}_{\tilde\ell_{ij}})^2=\lambda_iC_{ij} V_B
~~~\mbox{for}~~~~ i\neq j,\ee \be(m^{RR}_{\tilde\ell_{ij}})^2
=(m^{ij}_{\tilde R})^2-m_Z^2\cos{2\beta}\sin^2{\theta_W}.\ee In
analogy to the sneutrino mass term, we can arrange the parameters in
Eq. (34)-Eq.(36) so that $(m^{LL}_{\tilde\ell_{ij}})^2$,
$(m^{LR}_{\tilde\ell_{ii}})^2$ and $(m^{RR}_{\tilde\ell_{ij}})^2$
are zero for $i\neq j$ and there is no inter-generation lepton
number mixing in the slepton mass term. So doing, the slepton mass
matrix reduces to three individual mass matrices for three
generations \bea {\cal L}^{\tilde
\ell}_{mass}=\sum_{i=e,\mu,\tau}\left(\begin{array}{cc}\tilde\ell_{iL}^\ast
&
\tilde\ell^\ast_{iR}\end{array}\right)\left(\begin{array}{cc}(m^{LL}_{\tilde\ell_{ii}})^2
& (m^{LR}_{\tilde\ell_{ii}})^2\\(m^{LR}_{\tilde\ell_{ii}})^2 &
(m^{RR}_{\tilde\ell_{ii}})^2
\end{array}\right)\left(\begin{array}{c}\tilde\ell_{iL}\\ \tilde\ell_{iR}\end{array}\right). \eea
Since $\lambda_i V_B=m_i$, the off diagonal matrix element,
$(m^{LR}_{\tilde\ell_{ii}})^2$, can be written as\be
(m^{LR}_{\tilde\ell_{ii}})^2=m_i(\mu\tan\beta+C_{ii}).\ee Because
the masses of electron and muon are very small compared with the
sparticle mass scale, we ignore these off diagonal terms and
consider $\tilde\ell_{iL}$ and $\tilde\ell_{iR}$ as mass
eigenstates.

Inserting the transformation (18) and (28) in the interaction terms
(5)-(9) yields the explicit interactions in their mass basis:\be
{\cal L}^W_{int}=-\frac{g_2}{\sqrt
2}\sum_{i=e,\mu,\tau}\sum^6_{a=1}\Big(W^{-\mu}\bar
\ell_{iL}\gamma_\mu V_{ia} \nu_{a}+W^{+\mu}\bar
\nu_{a}V^{\ast}_{ia}\gamma_\mu \ell_{iL}\Big),\ee \be {\cal
L}^{\widetilde W^-}_{int}= -\frac{ig_2}{\sqrt
2}\sum_{i=e,\mu,\tau}\sum^2_{a=1}\bar \ell_{iL}~\widetilde
W^-~U^i_{1a}~\widetilde\nu_{ia}+\frac{g_2}{\sqrt
2}\sum_{i=e,\mu,\tau}\sum^4_{a=3}\bar \ell_{iL}~\widetilde
W^-~U^i_{3a}~\widetilde\nu_{ia}+H.C.,\ee \be {\cal L}^{\widetilde
W^0}_{int}=\frac{g_2i}{\sqrt 2}\sum_{i=e,\mu,\tau}\Big(\bar
\ell_{iL}~\widetilde W^0 ~\widetilde
\ell_{iL}-\widetilde\ell^\ast_{iL}~\overline{\widetilde
W^0}~\ell_{iL}\Big),\ee \be {\cal L}^{\widetilde
B}_{int}=\frac{g_1i}{\sqrt 2}\sum_{i=e,\mu,\tau}\Big(\bar
\ell_{iL}~\widetilde B ~\widetilde
\ell_{iL}-\widetilde\ell^\ast_{iL}~\overline{\widetilde
B}~\ell_{iL}\Big)+\sqrt 2 g_1i\sum_{i=e,\mu,\tau}\Big(\bar
\ell_{iR}~\widetilde B ~\widetilde
\ell_{iR}-\widetilde\ell^\ast_{iR}~\overline{\widetilde
B}~\ell_{iR}\Big),\ee \be {\cal L}^{\widetilde
h^-_B}_{int}=\sum_{i,j=e,\mu,\tau}\sum^2_{a=1}\frac{m_D^{ij}}{\sqrt
2 V_T}\bar
\ell_{iL}~\widetilde{h^-_B}~U^j_{2a}\widetilde\nu_{ja}+\sum_{i,j=e,\mu,\tau}\sum^4_{a=3}\frac{im_D^{ij}}{\sqrt
2 V_T}\bar
\ell_{iL}~\widetilde{h^-_B}~U^j_{4a}\widetilde\nu_{ja}+H.C..\ee

\section{The muonium-antimuonium oscillation in the EMSSM}

The lowest order Feynman diagrams accounting for muonium and
antimuonium mixing are displayed in Fig.1. Graphs (a), (b), (c) and
(d) are the non-SUSY contributions, which are mediated by Majorana
neutrinos and W boson. The other graphs all involve SUSY partners.
Graphs (e) and (f) are mediated by sneutrinos and winos, while graph
(g) and (h) are mediated by sneutrinos and higgsinos.
\begin{center}
\includegraphics[width=12cm]{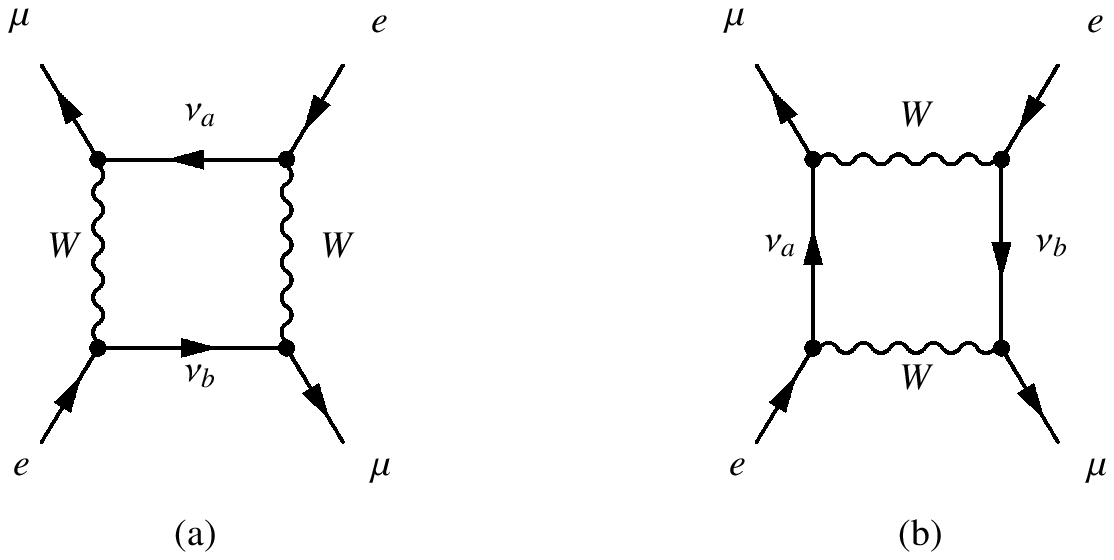}
\end{center}
\begin{center}
\includegraphics[width=12cm]{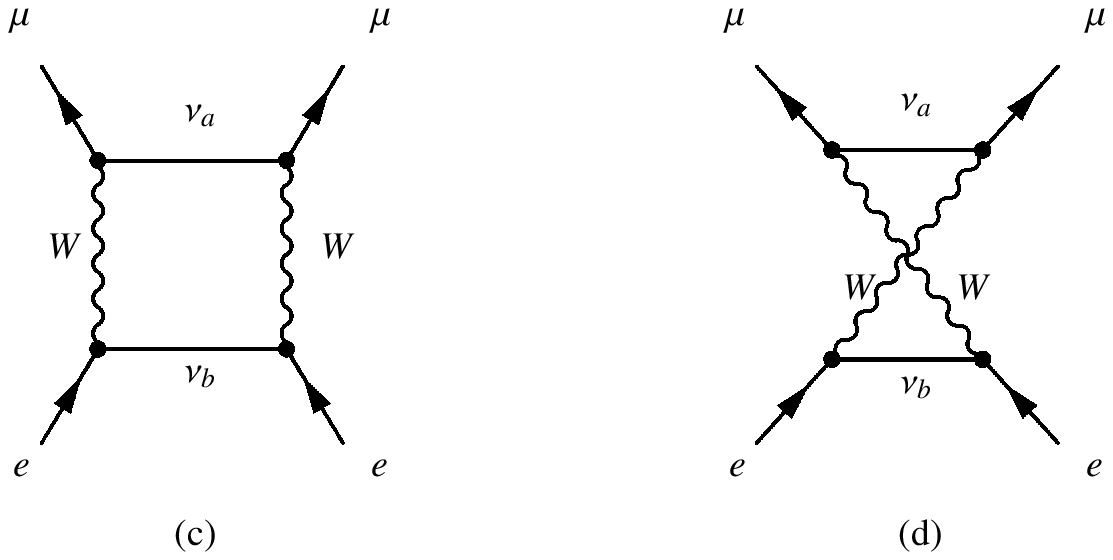}
\end{center}
\begin{center}
\includegraphics[width=12cm]{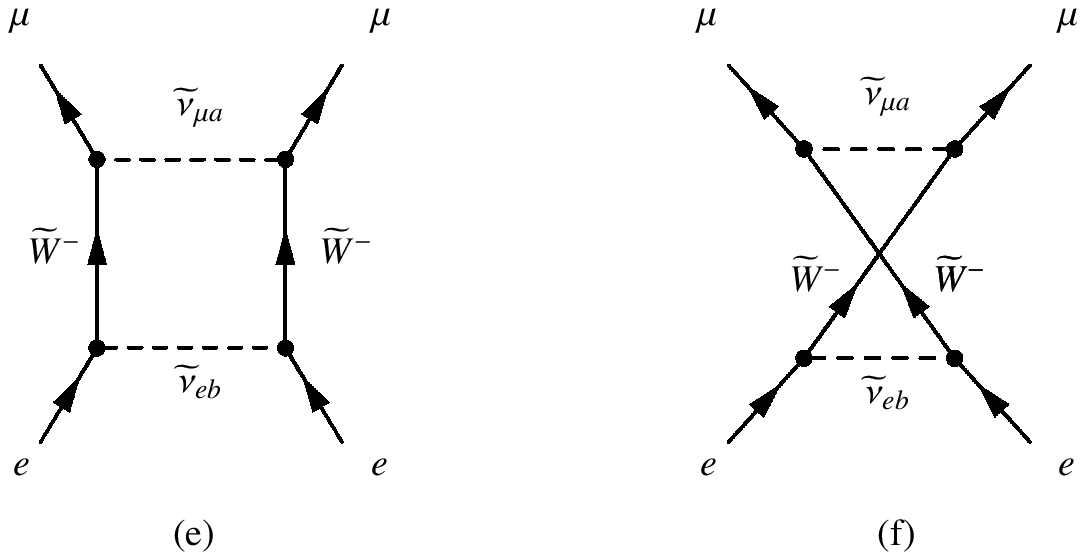}
\end{center}
\begin{center}
\includegraphics[width=12cm]{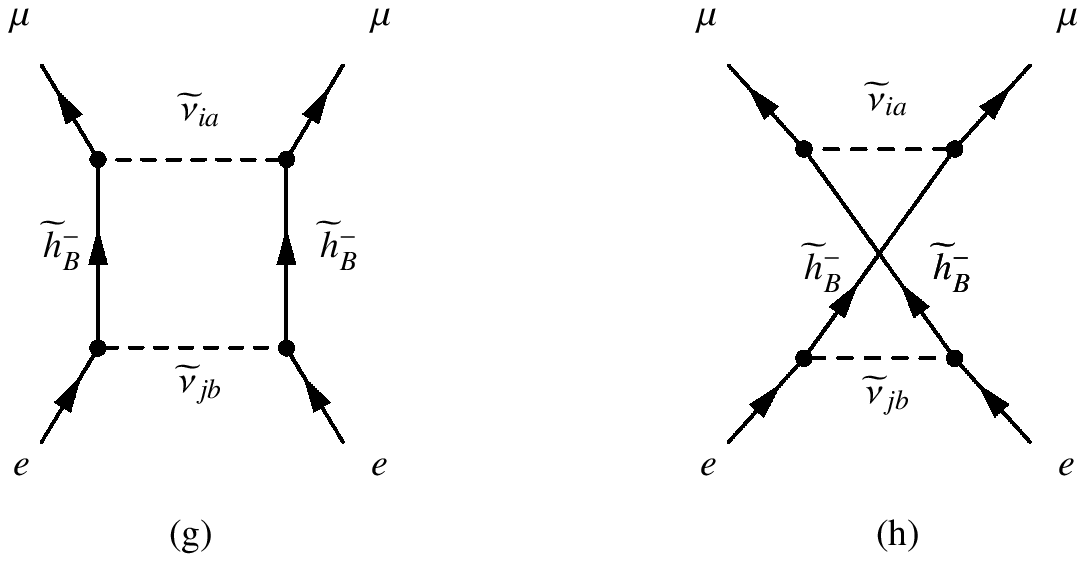}
\end{center}
\begin{center}
\figcaption{Feynman graphs contributing to the muonium-antimuonium
mixing.}
\end{center}
The T-matrix elements of graphs (a) and (b) are\cite{Liu}\bea
T_a=T_b =&&-\frac{g^4_2}{256\pi^2
M^2_W}[\bar\mu(3)\gamma_\mu(1-\gamma_5)e(2)]
[\bar\mu(4)\gamma^\mu(1-\gamma_5)e(1)]\cr
&&\Bigg[\sum^6_{a=1}(V_{\mu a}V^\ast_{ea})^2S(x_{\nu_a})
+\sum^6_{a,b=1;a\neq b}(V_{\mu a}V^\ast_{ea})(V_{\mu
b}V^\ast_{eb})T(x_{\nu_a},x_{\nu_b})\Bigg],\eea where
$\bar\mu(3)=\bar\mu(p_3,s_3)$ , $\bar\mu(4)=\bar\mu(p_4,s_4)$ ,
$e(1)=e(p_1,s_1)$ and $e(2)=e(p_2,s_2)$ are the spinors of the muons
and electrons and $x_{\nu_{a}}=\frac{m^2_{\nu_{a}}}{M^2_W}~,
~~~a=1,2,3,...6$. We define $S(x_{\nu_a})$ and
$T(x_{\nu_a},x_{\nu_b})$ as \be
S(x_A)=\frac{x^3-11x^2+4x}{4(1-x)^2}-\frac{3x^3}{2(1-x)^3}\ln
(x),\ee \be
T(x_A,x_B)=x_Ax_B\Big(\frac{R(x_A)-R(x_B)}{x_A-x_B}\Big)=T(x_B,x_A),\ee
with \be
R(x)=\frac{x^2-8x+4}{4(1-x)^2}\ln{(x)}-\frac{3}{4}\frac{1}{(1-x)}.\ee
The T-matrix elements of graphs (c) and (d) are\cite{Liu} \bea
T_c=T_d=&&\frac{g_2^4}{256\pi^2M^2_W}[\bar\mu(3)\gamma^\mu(1-\gamma_5)
e(2)][\bar\mu(4)\gamma_\mu(1-\gamma_5)e(1)]\cr
&&\cdot\Big[\sum^6_{a=1}(V_{\mu
a}V_{ea}^\ast)^2G(x_{\nu_a})+\sum^6_{a,b=1;a\neq b}(V_{\mu
a})^2(V_{eb}^\ast)^2K(x_{\nu_a},x_{\nu_b})\Big].\eea The functions
$G(x_{\nu_a})$ and $K(x_{\nu_a},x_{\nu_b})$ take the forms \bea
G(x_A)=\frac{(x_A-4)x_A}{(x_A-1)^2}+\frac{(x_A^3-3x_A^2+4x_A+4)x_A}{2(x_A-1)^3}\ln{x_A},
\eea\bea K(x_A,x_B)
=\sqrt{x_Ax_B}\frac{L(x_A,x_B)-L(x_B,x_A)}{x_A-x_B},\eea with\be
L(x_A,x_B)=\frac{4-x_Ax_B}{2(x_A-1)}+\frac{x_A(2x_B-x_Ax_B-4)}{2(x_A-1)^2}\ln
x_A.\ee The T-matrix elements of graphs (e) and (f) are\bea
T_e=T_f=&&-\frac{g_2^4}{1024\pi^2M^2_{\widetilde
W^-}}[\bar\mu(3)\gamma^\mu(1-\gamma_5)
e(2)][\bar\mu(4)\gamma_\mu(1-\gamma_5)e(1)]\cr
&&\cdot\Bigg(\sum^2_{a=1}\sum^2_{b=1}(U^\mu_{1a})^2(U^e_{1b})^2I(y_{\tilde\nu_{\mu
a}},y_{\tilde\nu_{eb}}
)-\sum^2_{a=1}\sum^4_{b=3}(U^\mu_{1a})^2(U^e_{3b})^2I(y_{\tilde\nu_{\mu
a}},y_{\tilde\nu_{eb}} )\cr
&&-\sum^4_{a=3}\sum^2_{b=1}(U^\mu_{3a})^2(U^e_{1b})^2I(y_{\tilde\nu_{\mu
a}},y_{\tilde\nu_{eb}}
)+\sum^4_{a=3}\sum^4_{b=3}(U^\mu_{3a})^2(U^e_{3b})^2I(y_{\tilde\nu_{\mu
a}},y_{\tilde\nu_{eb}} )\Bigg),\eea where \be
y_{\tilde\nu_{ia}}=\frac{m^2_{\tilde\nu_{ia}}}{M^2_{\tilde W^-}},\ee
\be I(x_1,x_2)=\frac{J(x_1)-J(x_2)}{x_1-x_2},\ee with \be
J(x)=\frac{x^2\ln{x}-x+1}{(x-1)^2}.\ee Finally, the T-matrix
elements of graphs (g) and (h) are\bea
T_g=T_h=&&-\sum_{i,j=e,\mu,\tau}\frac{(m^{\mu i}_Dm^{\mu
i\ast}_D)(m^{ej}_Dm^{ej\ast}_D)}{1024V^4_T\pi^2M^2_{\widetilde
h^-_B}}[\bar\mu(3)\gamma^\mu(1-\gamma_5)
e(2)][\bar\mu(4)\gamma_\mu(1-\gamma_5)e(1)]\cr
&&\cdot\Bigg(\sum^2_{a=1}\sum^2_{b=1}(U^i_{2a})^2(U^j_{2b})^2I(z_{\tilde\nu_{i
a}},z_{\tilde\nu_{jb}}
)-\sum^2_{a=1}\sum^4_{b=3}(U^i_{2a})^2(U^j_{4b})^2I(z_{\tilde\nu_{i
a}},z_{\tilde\nu_{jb}} )\cr
&&-\sum^4_{a=3}\sum^2_{b=1}(U^i_{4a})^2(U^j_{2b})^2I(z_{\tilde\nu_{i
a}},z_{\tilde\nu_{jb}}
)+\sum^4_{a=3}\sum^4_{b=3}(U^i_{4a})^2(U^j_{4b})^2I(z_{\tilde\nu_{i
a}},z_{\tilde\nu_{jb}} )\Bigg),\eea where \be
z_{\tilde\nu_{ia}}=\frac{m^2_{\tilde\nu_{ia}}}{M^2_{\widetilde
h^-_B}}.\ee

\section{The effective Lagrangian}

Combining all the T-matrix elements, we secure an effective
Lagrangian which can be cast as:\be {\cal L}_{eff}=\frac{G_{\bar
MM}}{\sqrt 2}[\bar \mu\gamma^\mu(1-\gamma_5)e][\bar
\mu\gamma_\mu(1-\gamma_5)e],\label{Leff}\ee where \bea \frac{G_{\bar
MM}}{\sqrt
2}=&&-\frac{g_2^4}{512\pi^2M^2_W}\cdot\Bigg(\sum^6_{a=1}(V_{\mu
a}V^\ast_{ea})^2\Big(S(x_{\nu_a})-G(x_{\nu_a})\Big)\cr
&&+\sum^6_{a,b=1;a\neq b}\Big((V_{\mu a}V^\ast_{ea})(V_{\mu
b}V^\ast_{eb})T(x_{\nu_a},x_{\nu_b})-(V_{\mu
a})^2(V_{eb}^\ast)^2K(x_{\nu_a},x_{\nu_b})\Big)\Bigg)\cr
&&-\frac{g_2^4}{2048\pi^2M^2_{\widetilde
W^-}}\cdot\Bigg(\sum^2_{a=1}\sum^2_{b=1}(U^\mu_{1a})^2(U^e_{1b})^2I(y_{\tilde\nu_{\mu
a}},y_{\tilde\nu_{eb}}
)-\sum^2_{a=1}\sum^4_{b=3}(U^\mu_{1a})^2(U^e_{3b})^2I(y_{\tilde\nu_{\mu
a}},y_{\tilde\nu_{eb}} )\cr
&&-\sum^4_{a=3}\sum^2_{b=1}(U^\mu_{3a})^2(U^e_{1b})^2I(y_{\tilde\nu_{\mu
a}},y_{\tilde\nu_{eb}}
)+\sum^4_{a=3}\sum^4_{b=3}(U^\mu_{3a})^2(U^e_{3b})^2I(y_{\tilde\nu_{\mu
a}},y_{\tilde\nu_{eb}} )\Bigg)\cr
&&-\sum_{i,j=e,\mu,\tau}\frac{(m^{\mu i}_Dm^{\mu
i\ast}_D)(m^{ej}_Dm^{ej\ast}_D)}{2048V^4_T\pi^2M^2_{\widetilde
h^-_B}}\cdot\Bigg(\sum^2_{a=1}\sum^2_{b=1}(U^i_{2a})^2(U^j_{2b})^2I(z_{\tilde\nu_{i
a}},z_{\tilde\nu_{jb}} )\cr
&&-\sum^2_{a=1}\sum^4_{b=3}(U^i_{2a})^2(U^j_{4b})^2I(z_{\tilde\nu_{i
a}},z_{\tilde\nu_{jb}}
)-\sum^4_{a=3}\sum^2_{b=1}(U^i_{4a})^2(U^j_{2b})^2I(z_{\tilde\nu_{i
a}},z_{\tilde\nu_{jb}} )\cr
&&+\sum^4_{a=3}\sum^4_{b=3}(U^i_{4a})^2(U^j_{4b})^2I(z_{\tilde\nu_{i
a}},z_{\tilde\nu_{jb}} )\Bigg).\eea


Muonium (antimuonium) is a nonrelativistic Coulombic bound state of
an electron and an anti-muon (positron and muon). The nontrivial
mixing between the muonium ( $|M >$ ) and antimuonium ($|\bar M >$)
states is encapsulated in the effective Lagrangian of Eq. (58) and
leads to the mass diagonal states given by the linear combinations
\be
|M_\pm>=\frac{1}{\sqrt{2(1+|\varepsilon|^2)}}[(1+\varepsilon)|M>\pm(1-\varepsilon)|\bar
M>],\ee where \be \varepsilon=\frac{\sqrt{\mathcal M_{M\bar
M}}-\sqrt{\mathcal M_{\bar MM}}}{\sqrt{\mathcal M_{M\bar
M}}+\sqrt{\mathcal M_{\bar MM}}}, \ee \be \mathcal M_{M\bar
M}=\frac{<M|-\int d^3r\mathcal L_{eff}|\bar M>}{\sqrt{<M|M><\bar
M|\bar M>}},~~~\mathcal M_{\bar MM}=\frac{<\bar M|-\int d^3r\mathcal
L_{eff}|M>}{\sqrt{<M|M><\bar M|\bar M>}}.\ee

Since the neutrino sector is expected, in general, to be CP
violating, these will be independent, complex matrix elements. If
the neutrino sector conserves CP, with $|M >$ and $|\bar M >$ CP
conjugate states, then $\mathcal M_{M\bar M}=\mathcal M_{\bar MM}$
and $\epsilon= 0$. In general, the magnitude of the mass splitting
between the two mass eigenstates is
 \bea |\Delta M|=2\left|Re\sqrt{\mathcal M_{M\bar M}\mathcal M_{\bar
 MM}}~~\right|.
\eea Since muonium and antimuonium are linear combinations of the
mass diagonal states, an initially prepared muonium or antimuonium
state will undergo oscillations into one another as a function of
time. The muonium-antimuonium oscillation time scale, $\tau_{\bar
MM}$, is given by \be\frac{1}{\tau_{\bar MM}}=|\Delta M|.\ee

We would like to evaluate $|\Delta M|$ in the nonrelativistic limit.
A nonrelativistic reduction of the effective Lagrangian of Eq.
(\ref{Leff}) produces the local, complex effective potential \bea
V_{eff}(\textbf{r})=8\frac{G_{\bar MM}}{\sqrt
2}\delta^3(\textbf{r}).\eea

Taking the muonium (anitmuonium) to be in their respective Coulombic
ground states, $\phi_{100}(\textbf{r})= \frac{1}{\sqrt{\pi a^3_{\bar
MM}}}e^{-r/a_{\bar MM}}$, where $a_{\bar
MM}=\frac{1}{m_{red}\alpha}$ is the muonium Bohr radius with
$m_{red}=\frac{m_em_\mu}{m_e+m_\mu}\simeq m_e$ the reduced mass of
muonium, it follows that \bea \frac{1}{\tau_{\bar MM}}\simeq &&2\int
d^3r\phi^\ast_{100}(\textbf{r})|ReV_{eff}(\textbf r)|\phi(\textbf
r)_{100}\cr=&& 16\frac{|ReG_{\bar MM}|}{\sqrt
2}|\phi_{100}(0)|^2=\frac{16}{\pi}\frac{|ReG_{\bar MM}|}{\sqrt
2}\frac{1}{a^3_{\bar MM}}.\eea Thus we secure an oscillation time
scale\be \frac{1}{\tau_{\bar MM}}\simeq
\frac{16}{\pi}\frac{|ReG_{\bar MM}|}{\sqrt 2} m_e^3\alpha^3.\ee

\section{Estimate of the effective coupling constant}

The present experimental limit\cite{Willmann} on the non-observation
of muonium-antimuonium oscillation translates into the bound \be
|ReG_{\bar MM}|\leq 3.0\times 10^{-3}G_F, \ee where
$G_F\simeq1.16\times10^{-5}GeV^{-2}$ is the Fermi scale. This limit
can then be used to construct some constraints on the parameters of
this model.

For simplicity, we set the neutrino Dirac mass matrix elements
$m^{ij}_D$ and the right-handed neutrino mass matrix elements
$M^{ii}_R$ to some common mass scales $m_D$ and $M_R$ respectively.
The light neutrino mass scale $m_\nu$ is of order $m^2_D/M_R$, while
the heavy neutrino mass scale is of order $M_R$.

Using these assumptions and taking into account the mixing matrices
approximations Eq.(22) and (32), we can simplify the effective
coupling constant Eq.(59) to a more manageable approximated form.
The contribution from graphs (a), (b), (c) and (d) in $G_{\bar MM}$
is $-\frac{g_2^4}{512\pi^2M^2_W}\cdot\Big(\sum^6_{a=1}(V_{\mu
a}V^\ast_{ea})^2\Big(S(x_{\nu_a})-G(x_{\nu_a})\Big)+\sum^6_{a,b=1;a\neq
b}\Big((V_{\nu a}V^\ast_{ea})(V_{\mu
b}V^\ast_{eb})T(x_{\nu_a},x_{\nu_b})-(V_{\mu
a})^2(V_{eb}^\ast)^2K(x_{\nu_a},x_{\nu_b})\Big)\Big)$. With the
limits of $m_{\nu_1}, ~m_{\nu_2}, ~m_{\nu_3}\sim \mathcal
{O}(\frac{m^2_D}{M_R})$ and $m_{\nu_4},~m_{\nu_5},~m_{\nu_6}\sim
\mathcal {O}(M_R)$, the contribution of graphs (a), (b), (c) and (d)
can be approximated as\bea &&\mbox{case
1:}~~~\frac{g_2^4}{512\pi^2M^2_W}\cdot\frac{m^4_D}{M_R^2M^2_W}\ln
\left(\frac{M_R M_W }{m^2_D}\right), ~~~~\mbox{a=1, 2, 3, ~~~b=1, 2,
3,}\cr &&\mbox{case
2:}~~~\frac{g_2^4}{512\pi^2M^2_W}\cdot\frac{m^8_D}{M_R^4M^4_W}\ln
\left(\frac{M_RM_W }{m^2_D}\right), ~~~~\mbox{a=1, 2, 3, ~~~b=4, 5,
6,}\cr &&\mbox{case
3:}~~~\frac{g_2^4}{512\pi^2M^2_W}\cdot\frac{m^4_D}{M^2_R M^2_W}\ln
\left(\frac{M_R}{M_W}\right), ~~~~\mbox{a=4, 5, 6, ~~~b=4, 5,
6}.\eea Taking $M_R$ as the largest mass parameter, the first case
and the third case are comparable, while the second one is
suppressed by a factor $m^4_D/(M^2_RM^2_W)$. Therefore, the
contribution from graphs (a), (b), (c) and (d) is roughly\bea
&&-\frac{g_2^4}{512\pi^2M^2_W}\cdot\Bigg(\sum^6_{a=1}(V_{\mu
a}V^\ast_{ea})^2\Big(S(x_{\nu_a})-G(x_{\nu_a})\Big)\cr
&&+\sum^6_{a,b=1;a\neq b}\Big((V_{\nu a}V^\ast_{ea})(V_{\mu
b}V^\ast_{eb})T(x_{\nu_a},x_{\nu_b})-(V_{\mu
a})^2(V_{eb}^\ast)^2K(x_{\nu_a},x_{\nu_b})\Big)\Bigg)\cr
&&\approx\frac{9\cdot g_2^4m^4_D }{256\pi^2M^2_R
M^4_W}\cdot\ln{\frac{M_R}{M_W}}.\eea The second term in Eq.(59) is
the contribution of graph (e) and (f), in which the function
$I(y_{\tilde\nu_{\mu a}},y_{\tilde\nu_{eb}})$ is a decreasing
function of $y_{\tilde\nu_{\mu a}}$ and $y_{\tilde\nu_{eb}}$. It
will be small for heavy seutrinos. To see this, we employ the
approximations Eq.(32) \bea &&
U^\mu_{11},~~U^e_{11},~~U^\mu_{33},~~U^e_{33}\sim \mathcal
{O}(1),\cr &&U^\mu_{12},~~U^e_{12},~~U^\mu_{34},~~U^e_{34}\sim
\mathcal {O} (\frac{m_D}{M_R}),\eea so that the terms involving
heavy sneutrinos will get an extra suppression from the mixing
matrix. Therefore, the contribution of graph (e) and (f) is
dominated by the term that only includes the light sneutrinos so
that \bea&&-\frac{g_2^4}{2048\pi^2M^2_{\widetilde
W^-}}\cdot\Bigg(\sum^2_{a=1}\sum^2_{b=1}(U^\mu_{1a})^2(U^e_{1b})^2I(y_{\tilde\nu_{\mu
a}},y_{\tilde\nu_{eb}}
)-\sum^2_{a=1}\sum^4_{b=3}(U^\mu_{1a})^2(U^e_{3b})^2I(y_{\tilde\nu_{\mu
a}},y_{\tilde\nu_{eb}} )\cr
&&-\sum^4_{a=3}\sum^2_{b=1}(U^\mu_{3a})^2(U^e_{1b})^2I(y_{\tilde\nu_{\mu
a}},y_{\tilde\nu_{eb}}
)+\sum^4_{a=3}\sum^4_{b=3}(U^\mu_{3a})^2(U^e_{3b})^2I(y_{\tilde\nu_{\mu
a}},y_{\tilde\nu_{eb}} )\Bigg)\cr&&\approx-\frac{g_2^4}{2048\pi^2
M^2_{\widetilde W^-}}\cdot\Big( I(y_{\tilde\nu_{\mu
1}},y_{\tilde\nu_{e1}})-I(y_{\tilde\nu_{\mu
1}},y_{\tilde\nu_{e3}})-I(y_{\tilde\nu_{\mu
3}},y_{\tilde\nu_{e1}})+I(y_{\tilde\nu_{\mu
3}},y_{\tilde\nu_{e3}})\Big).\eea Employing the squared-mass
difference between the two light sneutrinos in Eq.(30), the above
expression can be approximated as\bea&&-\frac{g_2^4}{2048\pi^2
M^2_{\widetilde W^-}}\cdot\Big( I(y_{\tilde\nu_{\mu
1}},y_{\tilde\nu_{e1}})-I(y_{\tilde\nu_{\mu
1}},y_{\tilde\nu_{e3}})-I(y_{\tilde\nu_{\mu
3}},y_{\tilde\nu_{e1}})+I(y_{\tilde\nu_{\mu
3}},y_{\tilde\nu_{e3}})\Big)\cr&&\approx-\frac{g_2^4}{2048\pi^2
M^2_{\widetilde W^-}}\cdot(y_{\tilde\nu_{\mu 1}}-y_{\tilde\nu_{\mu
3}})(y_{\tilde\nu_{e 1}}-y_{\tilde\nu_{e3}})\frac{\partial}{\partial
y_{\tilde\nu_{\mu 1}}}\frac{\partial}{\partial y_{\tilde\nu_{e
1}}}I(y_{\tilde\nu_{\mu 1}},y_{\tilde\nu_{e1}})\cr
&&\approx-\frac{g_2^4}{2048\pi^2 M^2_{\widetilde W^-}}\cdot
\frac{\Delta m^2_{\tilde\nu_{\mu}}}{M^2_{\widetilde
W^-}}\cdot\frac{\Delta m^2_{\tilde\nu_{e}}}{M^2_{\widetilde
W^-}}\cdot\frac{\partial}{\partial y_{\tilde\nu_{\mu
1}}}\frac{\partial}{\partial y_{\tilde\nu_{e 1}}}I(y_{\tilde\nu_{\mu
1}},y_{\tilde\nu_{e1}}),\eea where the squared-mass differences are
\bea &&\Delta
m^2_{\tilde\nu_{\mu}}=\frac{4(m^{\mu\mu}_D)^2(A_{\mu\mu}-\mu\cot\beta-B_{\mu\mu})}{M^{\mu\mu}_R},\cr
&&\Delta
m^2_{\tilde\nu_{e}}=\frac{4(m^{ee}_D)^2(A_{ee}-\mu\cot\beta-B_{ee})}{M^{ee}_R}.\eea
Assuming $A_{\mu\mu}=A_{ee}\equiv A$ and $B_{\mu\mu}=B_{ee}\equiv
B$, the squared-mass differences of light muon sneutrinos and light
electron sneutrinos are\bea \Delta m^2_{\tilde\nu_{\mu}}=\Delta
m^2_{\tilde\nu_{e}}\equiv\Delta
m^2_{\tilde\nu}=\frac{4m_D^2(A-\mu\cot\beta-B)}{M_R}\eea so that
Eq.(73) then simplifies to\be-\frac{g_2^4}{2048\pi^2 M^2_{\widetilde
W^-}}\cdot \frac{\Delta m^2_{\tilde\nu_{\mu}}}{M^2_{\widetilde
W^-}}\cdot\frac{\Delta m^2_{\tilde\nu_{e}}}{M^2_{\widetilde
W^-}}\cdot\frac{\partial}{\partial y_{\tilde\nu_{\mu
1}}}\frac{\partial}{\partial y_{\tilde\nu_{e 1}}}I(y_{\tilde\nu_{\mu
1}},y_{\tilde\nu_{e1}})\approx-\frac{g_2^4(\Delta
m^2_{\tilde\nu})^2}{2048\pi^2 M^6_{\widetilde
W^-}}\frac{\partial}{\partial y_{\tilde\nu_{\mu
1}}}\frac{\partial}{\partial y_{\tilde\nu_{e 1}}}I(y_{\tilde\nu_{\mu
1}},y_{\tilde\nu_{e1}}).\ee

The contribution from graph (g) and (h) is not dominated by the
terms involving only light sneutrinos even though $I(z_{\tilde\nu_{i
a}},z_{\tilde\nu_{jb}} )$ is a decreasing function of
$z_{\tilde\nu_{i a}}$ and $z_{\tilde\nu_{jb}}$, because these terms
get suppressed by the mixing matrix. The terms including only light
sneutrinos
$\tilde\nu_{i1},~~\tilde\nu_{i3},~~\tilde\nu_{j1},~~\tilde\nu_{j3}$
 roughtly gives \bea&&(U^i_{21})^2(U^j_{21})^2I(z_{\tilde\nu_{i
1}},z_{\tilde\nu_{j1}} )-(U^i_{21})^2(U^j_{43})^2I(z_{\tilde\nu_{i
1}},z_{\tilde\nu_{j3}} ) \cr
&&-(U^i_{43})^2(U^j_{21})^2I(z_{\tilde\nu_{i 3}},z_{\tilde\nu_{j1}}
)+(U^i_{43})^2(U^j_{43})^2I(z_{\tilde\nu_{i
3}},z_{\tilde\nu_{j3}})\cr\approx&&\frac{m_D^4}{M_R^4}\cdot
\frac{\Delta m^2_{\tilde\nu_{i}}\Delta
m^2_{\tilde\nu_{j}}}{M^4_{\tilde h^-_B}}\frac{\partial}{\partial
z_{\tilde\nu_{i 1}}}\frac{\partial}{\partial z_{\tilde\nu_{j
1}}}I(z_{\tilde\nu_{i 1}},z_{\tilde\nu_{j1}})\cr\sim&&\mathcal
{O}\left(\frac{1}{M^6_R} \right), \eea while the terms including one
light and one heavy sneutrino are roughly
\bea&&(U^i_{21})^2(U^j_{22})^2I(z_{\tilde\nu_{i
1}},z_{\tilde\nu_{j2}} )-(U^i_{21})^2(U^j_{44})^2I(z_{\tilde\nu_{i
1}},z_{\tilde\nu_{j4}} ) \cr
&&-(U^i_{43})^2(U^j_{22})^2I(z_{\tilde\nu_{i 3}},z_{\tilde\nu_{j2}}
)+(U^i_{43})^2(U^j_{44})^2I(z_{\tilde\nu_{i
3}},z_{\tilde\nu_{j4}})\cr
&&+(U^i_{22})^2(U^j_{21})^2I(z_{\tilde\nu_{i 2}},z_{\tilde\nu_{j1}}
)-(U^i_{22})^2(U^j_{43})^2I(z_{\tilde\nu_{i 2}},z_{\tilde\nu_{j3}} )
\cr &&-(U^i_{44})^2(U^j_{21})^2I(z_{\tilde\nu_{i
4}},z_{\tilde\nu_{j1}} )+(U^i_{44})^2(U^j_{43})^2I(z_{\tilde\nu_{i
4}},z_{\tilde\nu_{j3}})\cr\approx
&&\left(\frac{m_D}{M_R}\right)^2\cdot\frac{\Delta
M^2_{\tilde\nu_i}\cdot\Delta m^2_{\tilde \nu_j}}{M^4_{\tilde
h^-_B}}\cdot\left(\frac{M_{\tilde
h^-_B}}{M_R}\right)^4+\left(\frac{m_D}{M_R}\right)^2\cdot\frac{\Delta
M^2_{\tilde\nu_j}\cdot\Delta m^2_{\tilde \nu_i}}{M^4_{\tilde
h^-_B}}\cdot\left(\frac{M_{\tilde h^-_B}}{M_R}\right)^4 \cr
\sim&&\mathcal {O}\left(\frac{1}{M^6_R}\right),\eea where $\Delta
M^2_{\tilde\nu_i}$ is the heavy sneutrino squared-mass
difference\bea \Delta M^2_{\tilde\nu_i}=4B_{ii}M^{ii}_R. \eea Under
our approximations,\bea \Delta M^2_{\tilde\nu_i}\equiv\Delta
M^2_{\tilde\nu}=4BM_R. \eea The terms including two heavy sneutrinos
are roughly \bea&&(U^i_{22})^2(U^j_{22})^2I(z_{\tilde\nu_{i
2}},z_{\tilde\nu_{j2}} )-(U^i_{22})^2(U^j_{44})^2I(z_{\tilde\nu_{i
2}},z_{\tilde\nu_{j4}} ) \cr
&&-(U^i_{44})^2(U^j_{22})^2I(z_{\tilde\nu_{i 4}},z_{\tilde\nu_{j2}}
)+(U^i_{44})^2(U^j_{44})^2I(z_{\tilde\nu_{i
4}},z_{\tilde\nu_{j4}})\cr \approx && \frac{\Delta
M^2_{\tilde\nu_i}\cdot\Delta M^2_{\tilde\nu_j}}{M^4_{\tilde
h^-_B}}\cdot\frac{M^6_{\tilde
h^-_B}}{3M^6_R}\cr\approx&&\frac{(\Delta
M^2_{\tilde\nu})^2M^2_{\tilde h^-_B}}{3M^6_R}\cr\sim&&\mathcal
{O}\left(\frac{1}{M^4_R}\right). \eea Comparing the $M_R$
dependences of Eq.(77), (78) and (81), we see that the dominant term
is the one involving two heavy sneutrinos. Thus the contribution
from graph (e) and (f) can be approximated as \bea
&&-\sum_{i,j=e,\mu,\tau}\frac{(m^{\mu i}_Dm^{\mu
i\ast}_D)(m^{ej}_Dm^{ej\ast}_D)}{2048V^4_T\pi^2M^2_{\widetilde
h^-_B}}\cdot\Bigg(\sum^2_{a=1}\sum^2_{b=1}(U^i_{2a})^2(U^j_{2b})^2I(z_{\tilde\nu_{i
a}},z_{\tilde\nu_{jb}} )\cr
&&-\sum^2_{a=1}\sum^4_{b=3}(U^i_{2a})^2(U^j_{4b})^2I(z_{\tilde\nu_{i
a}},z_{\tilde\nu_{jb}}
)-\sum^4_{a=3}\sum^2_{b=1}(U^i_{4a})^2(U^j_{2b})^2I(z_{\tilde\nu_{i
a}},z_{\tilde\nu_{jb}} )\cr
&&+\sum^4_{a=3}\sum^4_{b=3}(U^i_{4a})^2(U^j_{4b})^2I(z_{\tilde\nu_{i
a}},z_{\tilde\nu_{jb}} )\Bigg)\cr\approx&&-\frac{3m^4_D(\Delta
M^2_{\tilde\nu})^2}{2048V^4_T\pi^2M^6_R}= -\frac{3g^4_2m^4_D(\Delta
M^2_{\tilde\nu})^2(1+\tan^2\beta)^2}{8192\pi^2M^6_RM_W^4\tan^4
\beta}.\eea

Combining the various contributions, the effective coupling constant
is thus roughly given by\bea |ReG_{\bar
MM}|\approx&&\Bigg|\frac{9g_2^4m^4_D}{256\pi^2M^2_RM^4_W}\cdot\ln{\frac{M_R}{M_W}}-\frac{g_2^4(\Delta
m^2_{\tilde\nu})^2}{2048\pi^2 M^6_{\widetilde
W^-}}\cdot\frac{\partial}{\partial y_{\tilde\nu_{\mu
1}}}\frac{\partial}{\partial y_{\tilde\nu_{e 1}}}I(y_{\tilde\nu_{\mu
1}},y_{\tilde\nu_{e1}})\cr && -\frac{3g^4_2m^4_D(\Delta
M^2_{\tilde\nu})^2(1+\tan^2\beta)^2}{8192\pi^2M^6_RM_W^4\tan^4
\beta}\Bigg|.\eea The first term in Eq.(83) is the dominant
contribution of  graph (a), (b), (c) and (d), which contains the
intermediate neutrino and W boson. This contribution appears in the
model in which all SUSY partners decoupled. The second term is the
dominant contribution of graph (e) and (f), in which wino and
sneutrino appear in the intermediate states. Finally, the third term
is the dominant contribution of graph (g) and (h), with intermediate
higgsino and sneutrinos lines. The second and third terms both
depend on the sneutrino mass splitting. This reflects the
intra-generation lepton number violating property of the
muonium-antimuonium oscillation process, because the sneutrino mass
splitting is generated by the $\Delta L=2$ operators in the
sneutrino mass matrix.  To compare the relative sizes of these three
terms, we use the current experimental limits of the neutrino and
sparticle masses.

The first terms in Eq.(83) has a factor $m^4_D/M^2_R$, which is
 the scale of the light neutrino mass square
$m_\nu^2\simeq m^4_D/M^2_R$ generated by see-saw mechanism. The
experimental constraints on neutrino masses are summerized in
reference\cite{neutrino mass} as \bea &&m_\nu(\mbox{electron
based})<2eV,\cr && m_\nu(\mbox{muon based})<0.19MeV,\cr &&
m_\nu(\mbox{tau based})<18.2MeV, \eea For instance, assuming \be
m_D\sim M_W,\ee \be m_\nu=\frac{m^2_D}{M_R}\sim1eV, \ee then the
right-handed neutrino mass scale is about \be M_R\sim10^{13}GeV.\ee
In this case, the first  terms in Eq.(83) is roughly \bea
\frac{9g_2^4m^4_D}{256\pi^2M^2_RM^4_W}\cdot\ln{\frac{M_R}{M_W}}=
\frac{9g_2^4m^2_\nu}{256\pi^2M^4_W}\cdot\ln{\frac{M_R}{M_W}}&\simeq&3.9\times10^{-28}GeV^{-2}.\eea

The second term in Eq.(83) depends on the light sneutrino
squared-mass difference $\Delta m^2_{\tilde\nu}$, which can be
written in terms of light sneutrino mass splitting $\Delta m_{\tilde
\nu}$ by \be \Delta m^2_{\tilde\nu}=2m_{\tilde\nu}\Delta m_{\tilde
\nu},\ee where $m_{\tilde\nu}$ is the mass scale of light
sneutrinos. So doing the second term in Eq.(83) can be written as
\bea -\frac{g_2^4(\Delta m^2_{\tilde\nu})^2}{2048\pi^2
M^6_{\widetilde W^-}}\cdot\frac{\partial}{\partial y_{\tilde\nu_{\mu
1}}}\frac{\partial}{\partial y_{\tilde\nu_{e 1}}}I(y_{\tilde\nu_{\mu
1}},y_{\tilde\nu_{e1}})=-\frac{g_2^4m^2_{\tilde\nu}(\Delta
m_{\tilde\nu})^2}{512\pi^2 M^6_{\widetilde
W^-}}\cdot\frac{\partial}{\partial y_{\tilde\nu_{\mu
1}}}\frac{\partial}{\partial y_{\tilde\nu_{e 1}}}I(y_{\tilde\nu_{\mu
1}},y_{\tilde\nu_{e1}}).\eea Reference \cite{Grossman} provides an
upper limit on the sneutrino mass splitting by calculating the
one-loop correction to the neutrino mass. Assuming that this
correction is no larger than the tree result gives \be \Delta
m_{\tilde\nu}\leq2\times10^3m_\nu.\ee  Relaxing this absence of fine
tuning constraint can substantially enhance the contribution of the
graph (e) and (f). Taking the sneutrino mass splitting to be of the
same order as sneutrino mass \be \Delta m_{\tilde\nu}\sim
m_{\tilde\nu},\ee and $m_{\tilde\nu_\mu}$, $m_{\tilde\nu_e}$ to be
the common mass scale $m_{\tilde\nu}$ gives\be y_{\tilde\nu_{\mu
1}}\sim y_{\tilde\nu_{e1}}\sim y_{\tilde\nu}=\frac{m^2_{\tilde
\nu}}{M^2_{\tilde W^-}}.\ee Eq.(90) can then be written as \be
-\frac{g_2^4m^2_{\tilde\nu}(\Delta m_{\tilde\nu})^2}{512\pi^2
M^6_{\widetilde W^-}}\cdot\frac{\partial}{\partial y_{\tilde\nu_{\mu
1}}}\frac{\partial}{\partial y_{\tilde\nu_{e 1}}}I(y_{\tilde\nu_{\mu
1}},y_{\tilde\nu_{e1}})\approx-\frac{g_2^4m^4_{\tilde\nu}}{512\pi^2
M^6_{\widetilde W^-}}\cdot\frac{\partial}{\partial y_{\tilde\nu_{\mu
1}}}\frac{\partial}{\partial y_{\tilde\nu_{e 1}}}I(y_{\tilde\nu_{\mu
1}},y_{\tilde\nu_{e1}})\Big|_{y_{\tilde\nu_{\mu
1}},y_{\tilde\nu_{e1}}= y_{\tilde\nu}}.\ee The function
$\frac{m^4_{\tilde\nu}}{M^6_{\widetilde W^-}}\cdot
\frac{\partial}{\partial y_{\tilde\nu_{\mu
1}}}\frac{\partial}{\partial y_{\tilde\nu_{e 1}}}I(y_{\tilde\nu_{\mu
1}},y_{\tilde\nu_{e1}})\Big|_{y_{\tilde\nu_{\mu
1}},y_{\tilde\nu_{e1}}= y_{\tilde\nu}}$ in  Eq.(94) is a decreasing
function of $M_{\widetilde W^-}$. In order to calculate the maximum
contribution of graph (e) and (f), we use the experimental lower
bound on $M_{\widetilde W^-}$ . Many experimental searches for
physics beyond the standard model have been conducted and provide
various constraints on SUSY parameter space. Tab.3 lists some of the
constraints\cite{Grivaz}.
\begin{center}
\begin{tabular}{|l c|l c|}\hline
Sparticle & lower limit [GeV] & Sparticle & lower limit
[GeV]\\\hline $\tilde \chi^\pm_1$ & 94.0 & $\tilde\mu_R$ & 94.0\\
$\widetilde\chi^0_1$ & 46.0 & $\tilde e_L$ & 107.0\\
$\widetilde\nu$ & 94.0 & $\tilde e_R$ & 73.0\\
 \hline
\end{tabular}
\end{center}
\begin{center}
\tabcaption{Experimental lower limits on SUSY particle masses}
\end{center}
Fixing the wino mass to its lower limit in Tab.3 \be M_{\tilde
W^-}=94.0GeV,\ee the contribution
$\frac{g_2^4m^4_{\tilde\nu}}{512\pi^2 M^6_{\widetilde
W^-}}\cdot\frac{\partial}{\partial y_{\tilde\nu_{\mu
1}}}\frac{\partial}{\partial y_{\tilde\nu_{e 1}}}I(y_{\tilde\nu_{\mu
1}},y_{\tilde\nu_{e1}})\Big|_{y_{\tilde\nu_{\mu
1}},y_{\tilde\nu_{e1}}= y_{\tilde\nu}}$ as a function of sneutrino
mass scale $m_{\tilde\nu}$ is shown in Fig.2.
\begin{center}
\includegraphics[width=12cm]{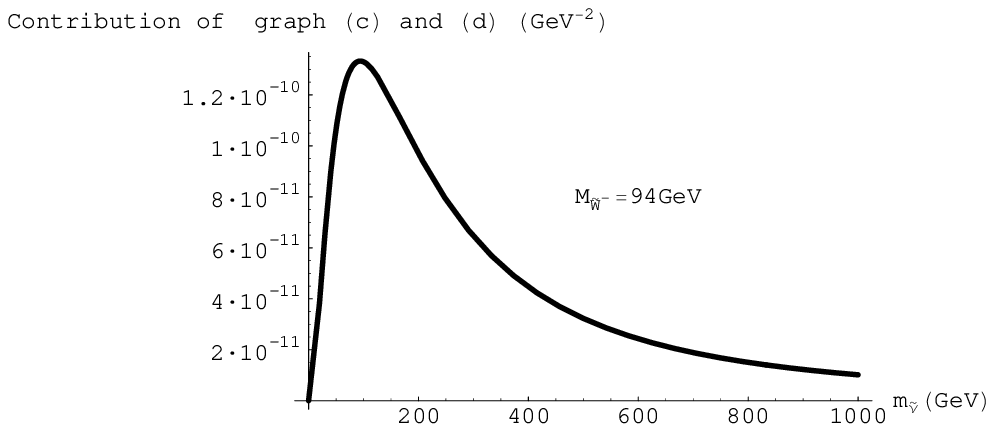}
\end{center}
\begin{center}
\figcaption{The contribution of graph (e) and (f) as a function of
$m_{\tilde\nu}$ when fixing the wino mass to its lower limit
$M_{\widetilde W^-}$.}
\end{center}
When $m_{\tilde\nu}=94.0GeV$, which is allowed by the experimental
limit in Tab.3, the contribution of graph (e) and (f) reaches its
maximum so that \be \frac{g_2^4m^4_{\tilde\nu}}{512\pi^2
M^6_{\widetilde W^-}}\cdot\frac{\partial}{\partial y_{\tilde\nu_{\mu
1}}}\frac{\partial}{\partial y_{\tilde\nu_{e 1}}}I(y_{\tilde\nu_{\mu
1}},y_{\tilde\nu_{e1}})\Big|_{y_{\tilde\nu_{\mu
1}},y_{\tilde\nu_{e1}}=
y_{\tilde\nu}}\leq1.3\times10^{-10}GeV^{-2}.\ee

Finally, the third term in Eq.(83) depends on the heavy sneutrino
squared-mass difference $\Delta M^2_{\tilde\nu}=2M_{\tilde\nu}\Delta
M_{\tilde\nu}=4BM_R$. Since we assume that $M_R$ is the largest mass
scale, $\Delta M_{\tilde\nu}$ can't be arbitrarily large. Taking
parameter $B$ one order of magnitude smaller than $M_R$, the heavy
sneutrino mass splitting is \be\Delta M_{\tilde\nu}\sim
\frac{M_R}{10}.\ee The contribution of graph (g) and (h) can then be
written as\bea \frac{3g^4_2m^4_D(\Delta
M^2_{\tilde\nu})^2(1+\tan^2\beta)^2}{8192\pi^2M^6_RM_W^4\tan^4
\beta}=\frac{3g^4_2m^2_\nu(1+\tan^2\beta)^2}{204800\pi^2M_W^4\tan^4
\beta}.\eea When $\tan\beta$ is very small the contribution can get
large and even reach the experimental limit Eq.(68). In this case,
the experimental limit of muonium-antimuonium oscillation provides
an inequality relating  $\tan\beta$ and $m_\nu$, which is given
by\be \frac{3g^4_2m^2_\nu(1+\tan^2\beta)^2}{204800\pi^2M_W^4\tan^4
\beta}\leq3.5\times10^{-8}GeV^{-2}.\ee This inequality translates
into a lower bound of $\tan\beta$ for different light neutrino
masses $m_\nu$: \bea
&&\tan\beta\geq6.6\times10^{-7},~~~~~~~~\mbox{if}~~~~m_\nu=1eV,\cr
&&\tan\beta\geq6.6\times10^{-8},~~~~~~~~\mbox{if}~~~~m_\nu=10^{-2}eV,\cr
&&\tan\beta\geq6.6\times10^{-9},~~~~~~~~\mbox{if}~~~~m_\nu=10^{-4}eV.\eea
The lower limit on $\tan\beta$ as a function of light neutrino mass
scale $m_\nu$ is shown in Fig.3.
\begin{center}
\includegraphics[width=12cm]{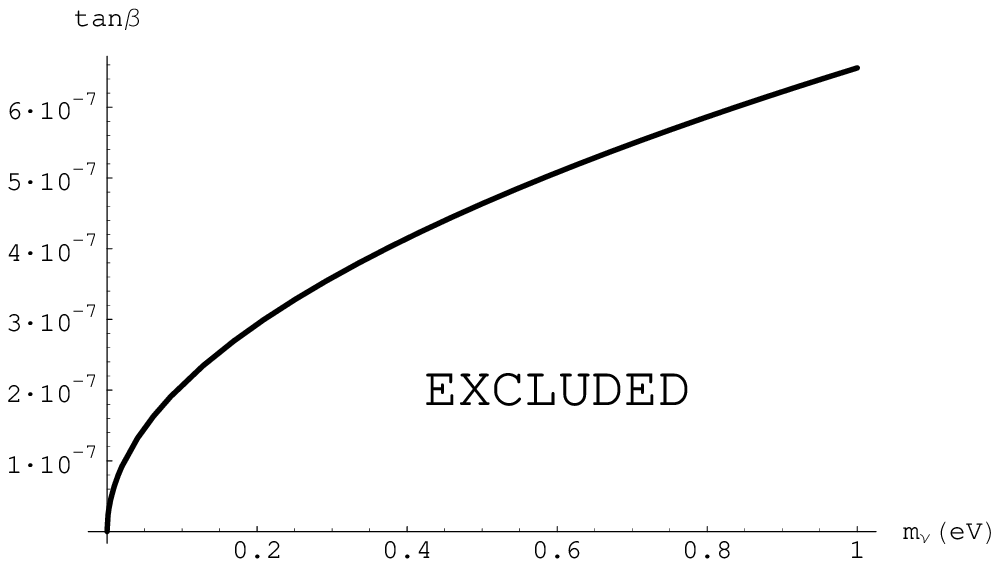}
\end{center}
\begin{center}
\figcaption{The lower limit on $\tan\beta$ as a function of light
neutrino mass scale $m_\nu$ provided by the muonium-antimuonium
oscillation experiment. The area above the curve is allowed by the
experiment results.}
\end{center} Notice that the ratio of the two Higgs VEVs $\tan\beta$
are related to the light neutrino masses in the above inequality,
although the graph (g) and (h) don't involve any neutrinos in the
intermediate states. This results since we are using a specific
model where the neutrino masses are generated by see-saw mechanism
$m_\nu\sim\mathcal {O}(m^2_D/M_R)$. The sneutrino mixing matrix is
approximated in term of $m_D/M_R$ and the heavy sneutrino masses are
also of order $M_R$. If we take $m_D$ to be of order $M_W$, the
heavy sneutrino masses $M_R$ in the contribution of graph (g) and
(h) can be expressed in term of the light neutrino mass scale
$m_\nu$. This explains the appearance of the parameter $m_\nu$ in
the inequality Eq.(99).

However, for non-infinitesimal values of $\tan\beta$, this
contribution is very small compared with the maximum of the second
term in Eq.(83). For instance, taking  the neutrino mass $m_\nu$ to
be $1eV$ and assuming $ \tan\beta\geq10^{-4}$,  the contribution of
graph (g) and (h) is \be
\frac{3g^4_2m^2_\nu(1+\tan^2\beta)^2}{204800\pi^2M_W^4\tan^4
\beta}\lesssim 6.5\times10^{-17}GeV^{-2}.\ee Thus, except for the
case of very small $\tan\beta$, the second term in Eq.(83) is the
dominant contribution for a wide range of the parameters and its
maximum is roughly two orders of magnitude below the sensitivity of
the current experiments.

\section{The constraints from the muon and electron anomalous
 magnetic moment experiments}

One has to be careful about other constraints on the model
parameters. Examples of such potential constraints come from the
measurements of the muon and electron anomalous magnetic moments.
The correction to the muon anomalous magnetic moment in the model
under consideration is found by calculating the one-loop graphs
shown in Fig.4.
\begin{center}
\includegraphics[width=12cm]{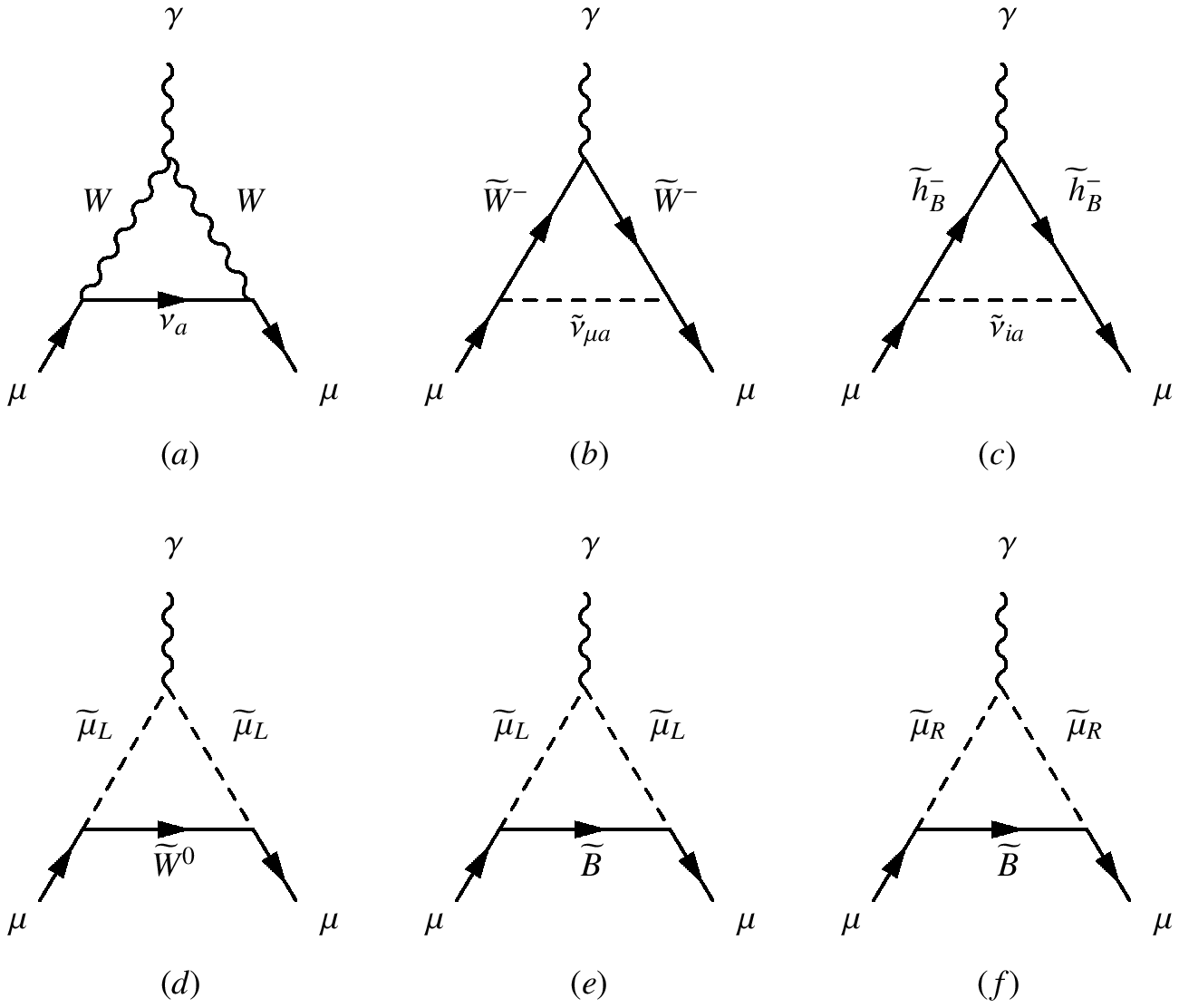}
\end{center}
\begin{center}
\figcaption{The Feynman graphs contributing to the muon anomalous
magnetic moment beyond the Standard Model.}
\end{center}

The muon anomalous magnetic moment contributed from the above graphs
is \bea
a^{BSM}_\mu=&&-\frac{g^2_2m^2_\mu}{16\pi^2M^2_W}\sum_{a=4,5,6}(V_{\mu
a}V^{\ast}_{\mu a})^2F^W(x_{\nu_{a}}) \cr
&&+\frac{g^2_2m^2_\mu}{32\pi^2M^2_{\tilde
W^-}}\left(\sum^2_{a=1}(U^\mu_{1a})^2F^C(y_{\tilde\nu_{\mu
a}})+\sum^4_{a=3}(U^\mu_{3a})^2F^C(y_{\tilde\nu_{\mu a}})\right)\cr
&&+\sum_{i=e,\mu,\tau}\frac{(m^{\mu i}_Dm^{\mu
i\ast}_D)m^2_\mu}{32\pi^2V_T M^2_{\tilde
h_B^-}}\left(\sum^2_{a=1}(U^i_{2a})^2F^C(z_{\tilde\nu_{i
a}})+\sum^4_{a=3}(U^i_{4a})^2F^C(z_{\tilde\nu_{i a}})\right)\cr
&&-\frac{g^2_2m^2_\mu}{32\pi^2M^2_{\tilde W^0}}F^N(s_{\tilde\mu_{
L}})-\frac{g^2_2m^2_\mu}{32\pi^2M^2_{\tilde B}}F^N(t_{\tilde
\mu_L})-\frac{g^2_1m^2_\mu}{8\pi^2M^2_{\tilde B}}F^N(t_{\tilde
\mu_R}),\eea where
$s_{\tilde\mu_a}=\frac{m^2_{\tilde\mu_a}}{M^2_{\tilde W^0}}$,
$t_{\tilde\mu_a}=\frac{m^2_{\tilde\mu_a}}{M^2_{\tilde B}}$, and \bea
F^W(x_{\nu_{a}})=\int^1_0dx\frac{-4x^2(1+x)-2x_\mu\cdot
x^2(x-1)-2x_{\nu_{a}}(2x-3x^2+x^3)}{x_\mu\cdot
x^2+(1-x_\mu)x+x_{\nu_{a}}(1-x)},\eea
\bea F^C(y_{\tilde\nu_{\mu a}})=\frac{2y_{\tilde\nu_{\mu
a}}^3-3y_{\tilde\nu_{\mu a}}^2(-1+2\ln{y_{\tilde\nu_{\mu
a}}})-6y_{\tilde\nu_{\mu a}}+1}{6(1-y_{\tilde\nu_{\mu a}})^4},\eea
\bea F^N(s_{\tilde\mu_{ a}})=\frac{s_{\tilde\mu_{
a}}^3-6s_{\tilde\mu_{ a}}^2+3s_{\tilde\mu_{ a}}+6s_{\tilde\mu_{
a}}\ln{s_{\tilde\mu_{ a}}}+2}{6(1-s_{\tilde\mu_{ a}})^4}.\eea

With assumption that $M_R$ is the largest mass scale, the dominant
contribution of the graphs in Fig.4 to $a^{BSM}_{\mu}$ is\bea
a^{BSM}_{\mu}\approx &&\frac{g^2_2 m^2_\mu m_D^2}{4\pi^2M^2_W
M^2_R}+\frac{g^2_2m^2_\mu}{32\pi^2M^2_{\widetilde
W^-}}\cdot\Big(F^C(y_{\tilde\nu_{\mu 1}})+F^C(y_{\tilde\nu_{\mu
3}})\Big)+\frac{g^2_2m^2_\mu
m^2_D(1+\tan^2\beta)}{32\pi^2M^2_WM^2_R\tan^2\beta}\cr
&&-\frac{g^2_2m^2_\mu}{32\pi^2M^2_{\tilde W^0}}F^N(s_{\tilde\mu_{
L}})-\frac{g^2_2m^2_\mu}{32\pi^2M^2_{\tilde B}}F^N(t_{\tilde
\mu_L})-\frac{g^2_1m^2_\mu}{8\pi^2M^2_{\tilde B}}F^N(t_{\tilde
\mu_R}).\eea The second and the last three terms are all decreasing
functions of slepton ,chargino and neutralino masses. We can use the
experimental bounds in Tab.3 to calculate the maximum values of
these terms yielding \bea\frac{g^2_2m^2_\mu}{32\pi^2M^2_{\widetilde
W^-}}\cdot\Big(F^C(y_{\tilde\nu_{\mu 1}})+F^C(y_{\tilde\nu_{\mu
3}})\Big) &\leq&2.5\times10^{-10},\cr
\frac{g^2_2m^2_\mu}{32\pi^2M^2_{\tilde W^0}}F^N(s_{\tilde\mu_{
L}})&\leq&1.9\times10^{-10},\cr
\frac{g^2_2m^2_\mu}{32\pi^2M^2_{\tilde B}}F^N(t_{\tilde
\mu_L})&\leq&1.9\times10^{-10},\cr\frac{g^2_1m^2_\mu}{8\pi^2M^2_{\tilde
B}}F^N(t_{\tilde \mu_R})&\leq&2.1\times10^{-10}.\eea The maxima of
 these terms are all about one order of magnitude smaller than the present experimental bound on the
contribution to $a_{\mu}=\frac{1}{2}(g-2)$ beyond the standard model
\cite{Kirill}: \be \delta
a_\mu=a^{exp}_\mu-a^{SM}_\mu=2\times10^{-9}.\ee The first and third
term both depend on the light neutrino mass scale $m_\nu$ and get
suppressed. For instance, using the assumptions Eq.(85) and Eq.(86),
the first and third terms are\bea\frac{g^2_2 m^2_\mu
m_D^2}{4\pi^2M^2_W M^2_R}=\frac{g^2_2 m^2_\mu m_\nu^2}{4\pi^2M^4_W
}&\approx&2.6\times10^{-30},\cr \frac{g^2_2m^2_\mu
m^2_D(1+\tan^2\beta)}{32\pi^2M^2_WM^2_R\tan^2\beta}=\frac{g^2_2m^2_\mu
m^2_\nu(1+\tan^2\beta)}{32\pi^2M^4_W\tan^2\beta}&\approx&
3.3\times10^{-31}\cdot\frac{1+\tan^2\beta}{\tan^2\beta}.\eea  The
first term is negligible compared with the terms in Eq.(107).
However, the third term can be large if $\tan\beta$ is very small.
Therefore, the experimental bound on the muon magnetic moment will
provide an inequality on $\tan\beta$ and $m_\nu$, wihch is given
by\bea\frac{g^2_2m^2_\mu
m^2_\nu(1+\tan^2\beta)}{32\pi^2M^4_W\tan^2\beta}\leq2\times10^{-9}.\eea
This inequality translates into a lower bound on $\tan\beta$ as a
function of the light neutrino mass scale $m_\nu$ as shown in Fig.5.
\begin{center}
\includegraphics[width=12cm]{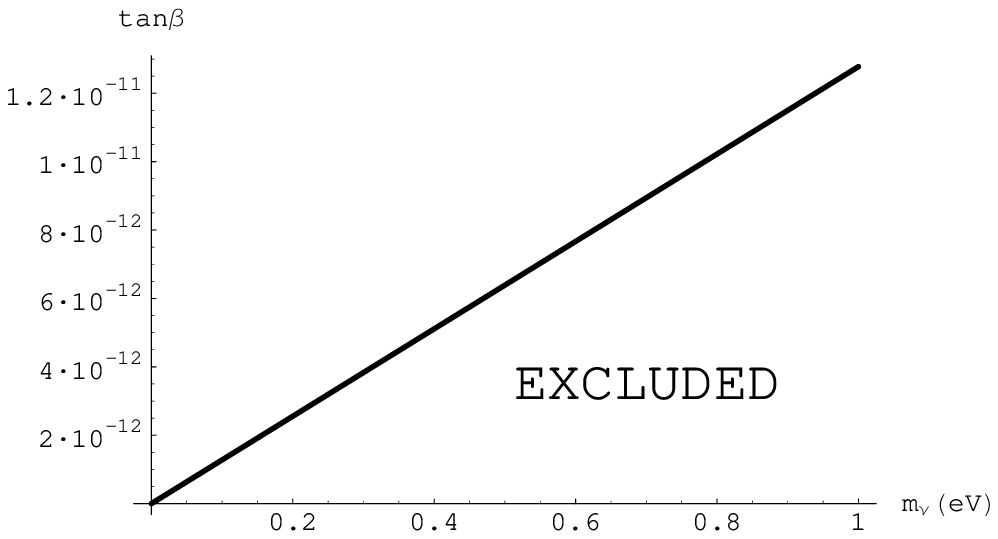}
\end{center}
\begin{center}
\figcaption{The lower limit on $\tan\beta$ as a function of light
neutrino mass scale $m_\nu$ provided by the muon anomalous magnetic
moment experiment. The area above the curve is allowed by the
experiment results.}
\end{center}

The electron anomalous magnetic moment beyond the Standard Model is
contributed by the six Feynman graphs displayed in Fig.6.
\begin{center}
\includegraphics[width=12cm]{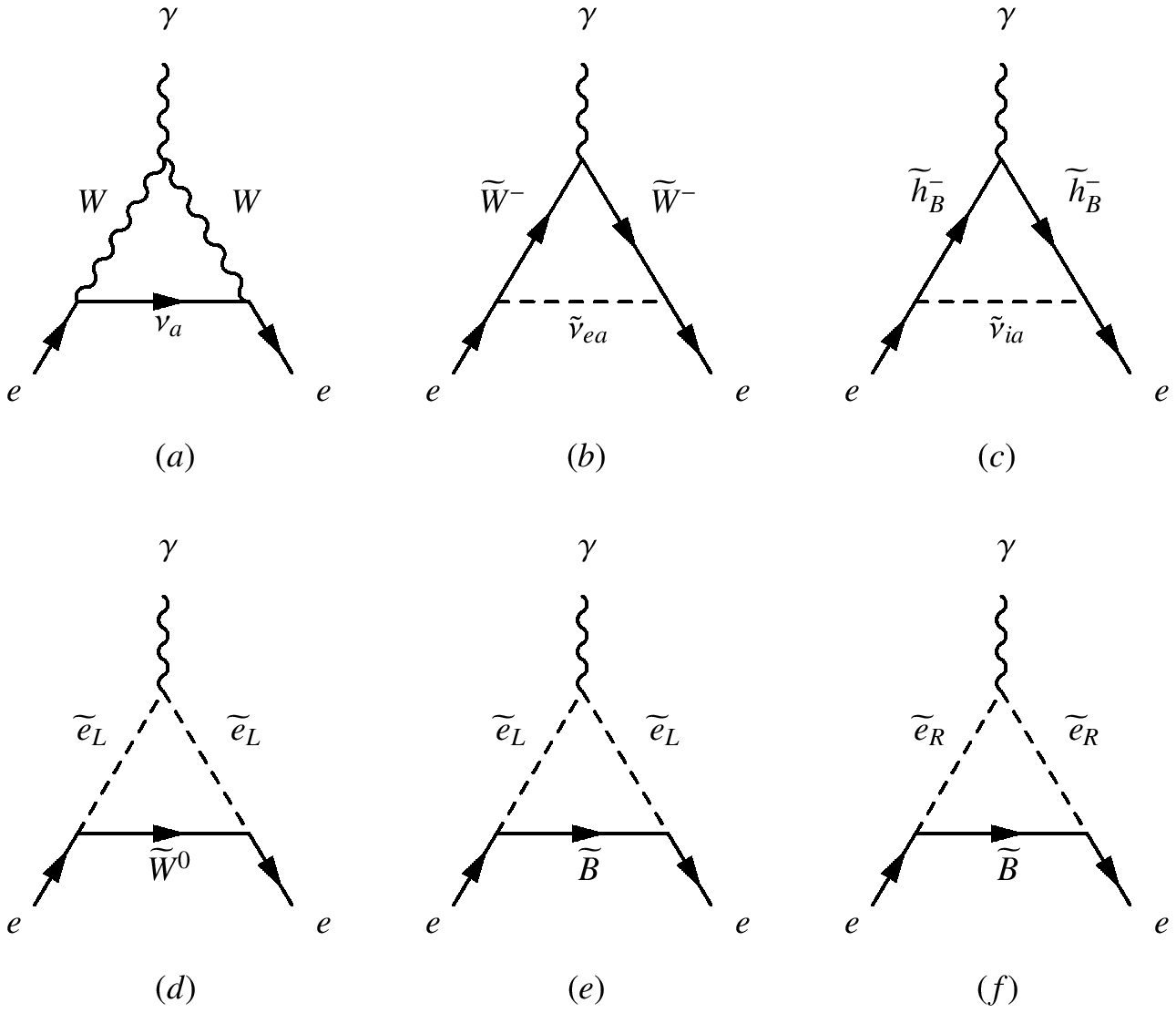}
\end{center}
\begin{center}
\figcaption{The Feynman graphs contributing to the electron
anomalous magnetic moment beyond the Standard Model.}
\end{center}

The experimental bound on the contribution to the electron anomalous
magnetic moment beyond Standard Model is\cite{John}\be \delta
a_e=a^{exp}_e-a^{SM}_e=1.4\times10^{-11}.\ee In analogy to the muon
case, this experimental limit will also generate an inequality
relation of $m_\nu$ and $\tan\beta$ given by\bea \frac{g^2_2m^2_e
m^2_\nu(1+\tan^2\beta)}{32\pi^2M^4_W\tan^2\beta}\leq1.4\times10^{-11}.\eea
This inequality is illustrated in Fig.7.\begin{center}
\includegraphics[width=12cm]{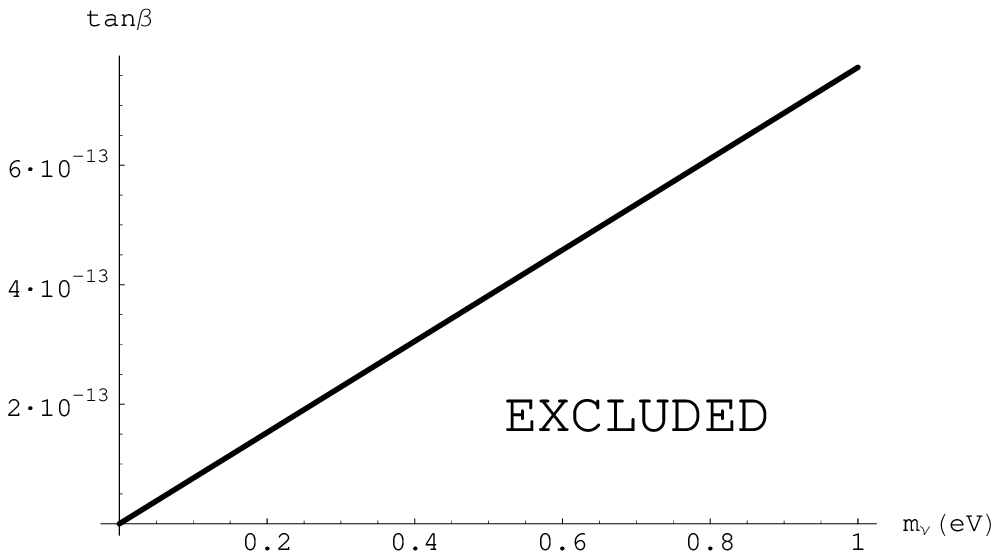}
\end{center}
\begin{center}
\figcaption{The lower limit on $\tan\beta$ as a function of light
neutrino mass scale $m_\nu$ provided by the electron anomalous
magnetic moment experiment. The area above the curve is allowed by
the experiment results.}
\end{center}

From Fig.2, Fig.5 and Fig.7 we see that for the model we are
considering, the muonium-antimuonium oscillation experiment gives a
more stringent constraint on $\tan\beta$ than the muon and electron
anomalous magnetic moment experiments.

\section{Conclusions}
We have calculated the effective coupling constant of the
muonium-antimuonium oscillation process in the Minimal
Supersymmetric Standard Model extended by inclusion of three
right-handed neutrino superfields where the required lepton flavor
violation has its origin in the Majorana property of the neutrino
and sneutrino mass eigenstates. For a wide range of the parameters,
the contribution of the graphs mediated by the sneutrino and winos,
$\widetilde W^-$, is dominant. The maximum of this contribution to
the effective coupling constant is roughly two orders of magnitude
below the sensitivity of current muonium-antimuonium oscillation
experiments. Moreover, these parameters are not constrained by the
$\mu\rightarrow e\gamma$ decay bounds, because the lowest order
contribution to the $\mu\rightarrow e\gamma$ process that involves
only sneutrino and winos in the intermediate states is forbidden in
the model. Finally, there is very limited possibility that the
contribution of the graphs mediated by sneutrinos and Higgsino
$\widetilde h^-_B$ is dominant if $\tan\beta$ is very small. In this
case, the contributions can even be large enough to reach the
present experimental bound. Therefore, the experimental bound can
provide an inequality on the model parameters,  which can be
translated into a lower bound on $\tan\beta$ as a function of light
neutrino mass scale $m_\nu$. The constraints from the muon and
electron anomalous magnetic moments were also investigated. For this
model, the muonium-antimuonium oscillation experiments give the most
stringent constraints on the parameters.

\begin{center}\section*{Acknowledgements}\end{center} It's a pleasure to thank Professor
S. T. Love, Professor T. E. Clark and Dr. Chi Xiong for useful
discussions.

\end{document}